\documentclass[11pt]{article}

\usepackage{fullpage}

\usepackage{amsmath,amssymb,amsthm,url,amsfonts,mathrsfs, bbm}
\usepackage{cancel}
\usepackage[bookmarks=true,pageanchor,colorlinks,linkcolor=purple,anchorcolor=purple,citecolor=purple,urlcolor=purple]{hyperref}
\usepackage{enumerate}
\usepackage{verbatim}
\usepackage{epsfig}
\usepackage{color}
\usepackage{graphics}
\usepackage{subfigure}
\usepackage{float}
\usepackage[font=small]{caption}
\usepackage{multirow}
\usepackage{anyfontsize}
\usepackage{cite}

\usepackage{times}

\usepackage{ifthen}
\newboolean{showcomments}
\setboolean{showcomments}{true}

\newcommand{\adam}[1]{\ifthenelse{\boolean{showcomments}}
{\textcolor{red}{(Adam says: \textit{#1}) }} {}}
\newcommand{\todo}[1]{\ifthenelse{\boolean{showcomments}}
{\textcolor{red}{(todo: \textit{#1}) }} {}}
\newcommand{\addref}[0]{\ifthenelse{\boolean{showcomments}}
{\textcolor{red}{(add ref) }} {}}
\newcommand{\addcite}[0]{\ifthenelse{\boolean{showcomments}}
{\textcolor{red}{(add cite) }} {}}
\newcommand{\addcites}[0]{\ifthenelse{\boolean{showcomments}}
{\textcolor{red}{(add cites) }} {}}
\newcommand{\bose}[1]{\ifthenelse{\boolean{showcomments}}
{\textcolor{red}{(Bose says: \textit{#1}) }}{}}
\newcommand{\desmond}[1]{\ifthenelse{\boolean{showcomments}}
{\textcolor{red}{(Desmond says: \textit{#1}) }}{}}

\newcommand{\beq}{\begin{equation}}
\newcommand{\eeq}{\end{equation}}
\newcommand{\beqa}{\begin{eqnarray}}
\newcommand{\eeqa}{\end{eqnarray}}
\newcommand{\beqan}{\begin{eqnarray*}}
\newcommand{\eeqan}{\end{eqnarray*}}

\newcommand{\vnorm}[1]{\left\|#1\right\|}

\newcommand{\diag}{\mathop{\mathrm{diag}}}

\newcommand{\Cset}{\mathbb{C}}

\newcommand{\Gset}{\mathbb{G}}

\newcommand{\Rset}{\mathbb{R}}

\newcommand{\Fcal}{{\cal F}}
\newcommand{\Gcal}{{\cal G}}

\newcommand{\Mcal}{{\cal M}}

\newcommand{\Pcal}{{\cal P}}

\newcommand{\Rcal}{{\cal R}}
\newcommand{\Scal}{{\cal S}}

\newcommand{\bone}{\mathbf{1}}

\renewcommand{\v}[1]{{\mathbf{#1}}}
\newcommand{\ve}{\varepsilon}

\newcommand{\sgn}{\text{sgn }}

\newcounter{l1}
\newcounter{l2}
\newcounter{l3}
\newcommand{\bdotlist}{\begin{list}{$\bullet$}{}}
\newcommand{\bboxlist}{\begin{list}{$\Box$}{}}
\newcommand{\bbboxlist}{\begin{list}{\raisebox{.005in}{{\tiny
$\blacksquare$ \ \ }}}{}}
\newcommand{\bdashlist}{\begin{list}{$-$}{} }
\newcommand{\blist}{\begin{list}{}{} }
\newcommand{\barablist}{\begin{list}{\arabic{l1}}{\usecounter{l1}}}
\newcommand{\balphlist}{\begin{list}{(\alph{l2})}{\usecounter{l2}}}
\newcommand{\bAlphlist}{\begin{list}{\Alph{l2}.}{\usecounter{l2}}}
\newcommand{\bdiamlist}{\begin{list}{$\diamond$}{}}
\newcommand{\bromalist}{\begin{list}{(\roman{l3})}{\usecounter{l3}}}

\newtheorem{theorem}{Theorem}
\newtheorem{lemma}{Lemma}
\newtheorem{proposition}{Proposition}

\newtheorem{corollary}{Corollary}

\renewcommand{\v}[1]{{\boldsymbol{#1}}}
\renewcommand{\Gset}{{\mathcal{G}}}
\renewcommand{\Cset}{{\mathcal{C}}}
\newcommand\Halmos{}

\usepackage{tikz}
\usetikzlibrary{snakes}
\usetikzlibrary{decorations.pathreplacing}
\usetikzlibrary{calc}

\graphicspath{{Plots/}}

\title{{\LARGE On the Role of a Market Maker in \\Networked Cournot Competition}\\ {\small Working paper}}

\author{Desmond Cai\thanks{D. Cai is with the Institute of High Performance Computing, Singapore. Email: desmond-cai@ihpc.a-star.edu.sg}
\ \ \ \ Subhonmesh Bose\thanks{S. Bose is with the Dept. of Electrical and Computer Engineering, University of Illinois at Urbana Champaign, Urbana, IL 61820, Email: boses@illinois.edu.} 
\ \ \ \ Adam Wierman\thanks{A. Wierman is with the Dept. of Computing and Mathematical Sciences, California Institute of Technology, Pasadena, CA 91125, Email: adamw@caltech.edu.}
}

\date{}

\begin{document}
\maketitle

\abstract{We study Cournot competition among firms in a networked marketplace that is centrally managed by a market maker.  In particular, we study a situation in which a market maker facilitates trade between geographically separate markets via a constrained transport network.  Our focus is on understanding the consequences of the design of the market maker and on providing tools for optimal design.  To that end we provide a characterization of the equilibrium outcomes of the game between the firms and the market maker.  Our results highlight that the equilibrium structure is impacted dramatically by the market maker's objective -- depending on the objective there may be a unique equilibrium, multiple equilibria, or no equilibria.  Further, the game may be a potential game (as in the case of classical Cournot competition) or not.  Beyond characterizing the equilibria of the game, we provide an approach for designing the market maker in order to optimize a design objective (e.g., social welfare) at the equilibrium of the game. Additionally, we use our results to explore the value of transport (trade) and the efficiency of the market maker (as compared to a single, aggregate market).}


\section{Introduction}

The ubiquity of networks in our world today has had a fundamental impact on modern marketplaces.  Classical models of competition often feature multiple firms operating in a single, isolated market; however power systems, the internet, transportation networks, infrastructure networks, and global supply chains are just a few of the places where varied and complex interconnections among participants are crucial to understanding and optimizing marketplaces.  Consequently, the study of competition in networked markets has emerged as an area with both rich theoretical challenges and important practical applications.

At this point, a wide variety of models for competition in networked markets have emerged across economics, operations research, and computer science.  The work in this literature focuses both on extensions of classical models of competition to networked settings, e.g., networked Bertrand competition \cite{guzman2011price, anshelevich2015price, chawla2008bertrand, acemoglu2009price} and networked Cournot competition \cite{bimpikis2014cournot,abolhassani2014network,ilkilic2009cournot}, and on models of specific applications where networked competition is fundamental, e.g., electricity markets \cite{neuhoff2005network, barquin08cournot, barquin2005cournot, yao2004computing, yao2007two, YaoAdlerOren2008, jing1999spatial}.

\subsection*{Intermediaries, market makers, and transport}

The complexity of networked marketplaces typically leads to (and often necessitates) the emergence of intermediaries.  A prominent illustration of this is financial markets, where central core banks intermediate trade between smaller periphery banks.  Similar examples are common in infrastructure networks: natural gas is traded through pipelines, which are managed by a Transmission System Operator (TSO), and transport in electricity markets is governed by an Independent System Operator (ISO).  One can view platforms in the sharing economy, e.g., Uber, as intermediaries between service providers and customers, and supply chains can be regarded as a form of intermediation in networked markets.

Intermediaries can play many roles in networked markets, from aggregation to risk mitigation to informational and beyond.  Our focus in this paper is on the role centralized intermediaries play with respect to transport and trade.  In particular, in many networked marketplaces participating firms depend on a centralized intermediary, a.k.a., market maker or platform, to provide transport of their goods between geographically distinct markets.

A particularly prominent example, which we use as the motivation throughout this paper, is \emph{electricity markets}. In these markets, the ISO solves a centralized dispatch problem by utilizing the offers/bids from the generators/retailers. This problem seeks to maximize some metric of social benefit subject to the operational constraints of the grid.
These operational constraints include physical laws that govern the flow of power in the network as well as safety constraints such as line capacity limits.
The payments are calculated based on locational marginal prices (LMP).
Therefore, the ISO plays a crucial role in matching the demand and supply of power within the confines of the grid and also define payments to the market participants. As an independent regulated entity, it further designs rules to limit the possible exercise of market power by the suppliers.

Beyond electricity markets, natural gas markets, and more generally, supply chains often have a similar structure where a market maker manages transport between geographically distributed markets.

Clearly, the design of the market maker in such situations is crucial to the efficiency of the marketplace.  By facilitating trade, the market maker is providing a crucial opportunity for increased efficiency.  However, constraints inherent to the transport network can make it difficult to realize this potential. As an example, network constraints can give rise to hidden monopolies, where even a small firm can exhibit market dominance because of its position in the network 

The dangers of such hidden monopolies are especially salient (and the corresponding efficiency loss is especially large) in the case of electricity markets, since power flows cannot be controlled in an end-to-end manner due to Kirchhoff's laws.  Even though California's electricity crisis is long past, examples of generators attempting to exploit this sort of market power are still common today, e.g., JP Morgan was fined \$410 million for market manipulations in California and the midwest from September 2010 to November 2012 \cite{jpmorgan}, and are expected to become more prominent as the penetration of renewable energy grows \cite{ruhi2017opportunities}.

\subsection*{Contributions of this paper.} \emph{Our goal in this paper is to provide insight into the design (and regulation) of market makers that govern transport in networked marketplaces.} In particular, we study a model of networked Cournot competition in which transport between geographically distinct markets is governed by a market maker (market operator) and subject to network flow constraints.  Our results focus on the impact the design of the market maker has on the equilibrium outcomes of the game between firms and the market maker.

Our first contribution is the model itself.  We introduce a general, parameterized model of a market maker (Section \ref{sec:model}) in a centrally managed networked Cournot competition. This model generalizes the networked market models used in the electricity markets community (see \cite{ventosa2005electricity} for a survey).  In our model, each market contains multiple firms competing locally in a Cournot competition and there is a simultaneous move game between the market maker and the firms. The model is distinctive in that the market maker controls transport, acting as an intermediary between markets by buying from some markets and selling to other markets and using its network to transport the goods between markets subject to the constraints of the network. Motivated by work in electricity markets, we focus on a specific form of market maker which clears the market by maximizing a payoff function that is parameterized by the tradeoff between the benefit to each of the three key parties -- the consumer, the producer, and the market maker itself.

Our second contribution is the characterization of the equilibria structure as a function of the design parameters of the market maker (Section \ref{sec:NEchar}).  Our main result (Theorem \ref{thm:main}) highlights a wide variety of behaviors -- depending on the design of the market maker, there may be a unique equilibrium, multiple equilibria, or no equilibria.  Further, when equilibria do exist, the game may form a weighted potential game or not depending on the design choice.  Beyond characterizing existence of equilibria, in the case of linear costs, homogeneous demands, and an unconstrained network, we are able to explicitly characterize the unique equilibrium outcome as a function of the market maker design.  This allows us to perform a more detailed study of the impact of the market maker.  For example, the characterization highlights that the total production by all firms is independent of the design of the market maker (in this setting), but that the relative production of the firms may vary dramatically depending on the design of the market maker.  Additionally, the characterization allows us to provide results highlighting the value of the trade provided by the market maker as well as the efficiency of the market maker (i.e., how close the outcomes of the game are to the outcomes of a single, aggregate Cournot market) as a function of the market maker design.

Our third contribution focuses on the design of the market maker.  In particular, we show how to (approximately) optimally design the market maker payoff so as to maximize a desired social/regulatory objective, e.g., social welfare, (Section \ref{sec:design}). These results provide insight into how the market maker, e.g., the ISOs in the case of the electricity market, may adjust their clearing rules in order to improve social welfare.  Our primary tool is the characterization of the equilibria provided in Section \ref{sec:NEchar}.  Then, we utilize the sum of squares (SOS) relaxation framework to judge the quality of our approximately optimal design choice. The results highlight the, perhaps counterintuitive, observation that if the market maker intends to optimize social welfare, it should not use social welfare as the objective in clearing the market. The intuition is that the market maker can exploit its commitment strategy (to the market clearing rule) to change the Nash equilibrium to its advantage. We further illustrate our proposed approach to market maker design on a stylized example that represents a caricature of the California electricity market. Our results underscore the importance of careful design.

\subsection*{Related literature.}

Models of competition in networked settings have received considerable attention in recent years.  These models come in various forms, including networked Bertrand competition, e.g., \cite{guzman2011price, anshelevich2015price, chawla2008bertrand, acemoglu2009price}, networked Cournot competition, e.g., \cite{bimpikis2014cournot,abolhassani2014network,ilkilic2009cournot}, and various other non-cooperative bargaining games where agents can trade via bilateral contracts and a network determines the set of feasible trades, e.g., \cite{Elliott2015,CondorelliGaleotti2012,Nava2015,AbreuManea2012,Manea2011}.

Our paper fits into the emerging literature on networked Cournot competition; however our focus and model differ considerably from existing work. In particular, beginning with \cite{BulowGeanakoplosKlemperer1985} and continuing through \cite{ilkilic2009cournot,bimpikis2014cournot,abolhassani2014network}, the literature on networked Cournot competition has focused on models where the network structure emerges as a result of firms having a fixed, limited set of markets in which they can participate and participation in these markets is unconstrained and independent of the actions of other firms.  In contrast, in our model the network constrains flows between markets, and so there are coupled participation constraints for the firms.  Further, the literature on networked Cournot competition has focused on situations where firms operate independently, without governance, while we focus on situations where transport across markets is managed by a market maker.

The line of work that is most relevant to the questions studied in this paper comes from the electricity market literature, where versions of Cournot competition subject to network constraints have been studied for nearly two decades, see \cite{ventosa2005electricity} for a survey.  In this setting Cournot models often provide good explanations for observed price variations~\cite{willems2009cournot}, and so are quite popular. For example, Cournot models have been applied to perform detailed studies of electricity markets in the US \cite{BorensteinBushnell1999},  Scandinavia \cite{AnderssonBergman1995}, Spain \cite{AlbaOtero-NovasMeseguerBatlle1999,RamosVentosaRivier1999}, and New Zealand \cite{Scott1998,ScottRead1996}, among others.

Due to the importance of the ISO in electricity markets, papers within this literature often include a model of a market maker, e.g.,  \cite{willems2002modeling,yao2004computing,metzler2003nash,jing1999spatial,Hobbs2001,YaoAdlerOren2008, bose2014cdc}. Our model builds on and generalizes the models for market makers and network competition in this literature.  More specifically, with rare exception, these papers focus on a market maker that is regulated to maximize social welfare, and thus do not explore the impact of differing market maker payoffs, nor how to design the market maker to optimize a particular social objective.  Further, these papers focus exclusively on detailed models of power flows, and thus do not apply to more general network models, such as classical flow models, which are relevant to other applications. Our results, on the other hand, apply to networks with general linear constraints, including both linearized power flow constraints and classical network flow constraints.

Note that, due to the operational constraints of electricity markets, the power of the market maker is limited, and thus general market mechanisms are not feasible.  See the survey of \cite{ventosa2005electricity} for a discussion of why.  In this paper we follow the assumptions in this line of work, which means that prescriptions from this paper can be used to improve efficiency (when the market maker is the social planner) with limited changes to the marketplace design. However, more generally, if one was willing to radically change the marketplace design it would be possible to use techniques from mechanism design theory to design market makers for networked markets, e.g., see \cite{borgers2015introduction, garg2008foundations, shoham2008multiagent}.

To the best of our knowledge, this is the first paper to focus on understanding the impact of, and how to optimally design, a market maker that governs transport in a networked marketplace. This paper builds on our preliminary work described in, \cite{bose2014cdc}, which is a short illustration of the contrast between three particular market clearing rules in an example two-node network. The work in the current paper considers a more general model of network constraints, studies general networks, and most importantly characterizes the equilibrium structure (existence, uniqueness, potential game) for a general parameterized class of market maker designs. In addition, it also provides a systematic approach for optimizing the market maker's objective.


\section{Model}
\label{sec:model}

Our focus is a marketplace where a constrained transport network, operated by a market maker, connects firms and markets.  Specifically, we consider an economy dealing in a single commodity that is comprised of a set of markets $\Mcal$, a set of firms $\Fcal$, and a market maker who facilitates transport of the commodity between the markets. Within this setting, we study Cournot competition over the networked markets, considering a static game of complete information among the firms and the market maker.

Each firm $f \in \Fcal$ supplies to exactly one market\footnote{In our motivating example of electricity markets, generators supply power only at a fixed location in the network. We model that spatial fixity of suppliers by allowing each firm to compete only in a single market, as opposed to the models considered in \cite{bimpikis2014cournot,abolhassani2014network,ilkilic2009cournot}.}, denoted by $\Mcal(f)$. Let $\Fcal(m)$ denote the set of firms that supply to market $m \in \Mcal$. Denote the supply of firm $f \in \Fcal$ to market $\Mcal(f)$ by $q_f \in \Rset_+$, and let $ \v{q} := \left( q_f, f\in\Fcal \right) \in \Rset^{| \Fcal |}_+ $
denote the vector of supplies of all firms in $\Fcal$. Additionally, for each $f \in \Fcal$, let $\v{q}_{-f}$ denote the vector of supplies of all firms in $\Fcal$, except $f$. The cost incurred by firm $f \in \Fcal$ for producing $q_f \in \Rset_+$ is $c_f(q_f)$. Assume $c_f : \Rset_+ \to \Rset_+$ is nondecreasing, convex, twice continuously differentiable, and $c_f(0) = 0$.

Crucially, the production of each firm in our model can be reallocated to other markets by a market maker that controls a constrained transport network.  We consider a single market maker that facilitates transport of the commodity between markets. The market maker can procure supply from one market and transport it to a different market, subject to network constraints. Denote the quantity supplied by the market maker to market $m \in \Mcal$ by $r_m$. Our convention is that $r_m \geq 0$ ($r_m < 0$) denotes a net supply (net demand) of the commodity by the market maker in market $m$. For convenience, let $ \v{r} := \left( r_m, \ m\in \Mcal \right) \in \Rset^{| \Mcal |}$ denote the vector of supplies by the market maker. Since the market maker only transports the commodity, the market maker neither consumes nor produces. So, we have $\bone^\top \v{r} = 0$, where $\bone$ is a vector of ones with dimension $| \Mcal|$.\footnote{We recognize that, in some cases, the market maker may have an incentive to dispose off some of its purchases. We can model such behavior by replacing the constraint $\bone^\top \v{r} = 0$ with $\bone^\top \v{r} \leq 0$. Most of our results continue to hold with the latter constraint. However, our motivating application of electricity markets does not feature disposal; hence, we assume $\bone^\top \v{r} = 0$ throughout.}

The reallocation of supply by the market maker, $\v{r}$, is subject to the flow constraints of the network.  We model these constraints by restricting $\v{r}$ to a polyhedral set
$ \Pcal := \left\{ \v{r} : \v{A} \v{r} \leq \v{b} \right\}  \subseteq \Rset^{| \Mcal |},$
where $\v{A}$ and $\v{b}$ define the half-spaces of $\Pcal$.  This formulation can capture constraints in traditional flow networks, as well as power flow constraints arising from linearized Kirchoff's laws and line limits. We remark that our results can be generalized to $\Pcal$ being a general convex semi-algebraic set with nonempty interior.  

In representing the flow constraints we have not allowed for storage.  Lack of meaningful inventory in the power system has prompted this decision.  However, if one were to desire to include storage within our model, participation of competitive storage will only alter the intercepts of the inverse demand curves. But, the case with strategic storage ownership leads to a dynamic game because of inter-temporal considerations.  This is an interesting topic for future work.  

The price at each market in the network is dependent on both the production of the firms and the reallocation performed by the market maker.  As is traditional when studying Cournot competition, we focus on the case of linear inverse demand functions.  In particular, assume that the price $p_m$ in each market $m \in \Mcal$ has the form
\begin{align}
\label{eq:dFunc}
p_m(d_m) := \alpha_m - \beta_m d_m
\end{align}
for some $\alpha_m, \beta_m > 0$. Here, $d_m$ is the aggregate demand in market $m$. Importantly, the aggregate demand in each market is determined by both the actions of the firms and the market maker, i.e., $d_m = r_m + \sum_{f \in \Fcal(m)} q_f$.

The payoff of firm $f \in \Fcal$ is given by its profit, defined as
\begin{align}
\pi_{f}(\v{q}, \v{r}) := q_{f} \cdot p_{\Mcal(f)}\left(r_{\Mcal(f)} + \sum_{f' \in \Fcal(\Mcal(f))} q_{f'} \right) - c_f(q_f). \label{eq:defpif}
\end{align}
Thus, firm $f$ maximizes $\pi_{f}(\v{q}, \v{r})$ over $q_f \in \Rset_+$, given $(\v{q}_{-f}, \v{r})$.

For the market maker, the payoff function is a design choice.  In many regulated settings, e.g., electricity markets, it is common for the market maker to optimize some metric of social benefit.  Our goal in this paper is to explore the impact of the market maker payoff functions, and so we focus on a broad parameterized class of maker maker payoff functions defined as follows.  Given $\v{q}$, the market maker maximizes $\Pi(\v{q}, \v{r}; \v{\theta})$ over $\v{r} \in \Pcal$ and $\bone^\top \v{r} = 0$,  where
\begin{align}
\Pi(\v{q}, \v{r}; \v{\theta})
:= \sum_{m \in \Mcal} \left[ \theta_C  \cdot \mathsf{CS}_m(\v{q}, \v{r}) + \theta_P\cdot\mathsf{PS}_m(\v{q}, \v{r}) + \theta_M\cdot\mathsf{MS}_m(\v{q}, \v{r}) \right].
\label{eq:defPi}
\end{align}
In $\Pi(\v{q}, \v{r}; \v{\theta})$, the design parameter $\v{\theta} := \left(\theta_C, \theta_P, \theta_M \right)^\top \in \Rset^3_+$ allows the designer to weigh the importance of the following terms, for each $m \in \Mcal$:\footnote{The notation $\v{\theta}\in\Rset_+^3$ should be understood to mean that $\v{\theta} \gneq \v{0}$ since the market maker's payoff becomes zero trivially when $\v{\theta} = \v{0}$.}
\begin{align*}
\mathsf{CS}_m(\v{q}, \v{r}) &:= \displaystyle\int_0^{r_{m} + \sum_{f \in \Fcal(m)} q_{f}} p_m(w_m) \, dw_m - \left(r_{m} + \sum_{f \in \Fcal(m)} q_{f} \right) \cdot p_m\left(r_{m} + \sum_{f \in \Fcal(m)} q_{f} \right);\\
\mathsf{PS}_m(\v{q}, \v{r}) &:=  \left(\sum_{f \in \Fcal(m)} q_f \right) \cdot p_{m}\left(r_{m} + \sum_{f \in \Fcal(m)} q_{f} \right)   - \sum_{f \in \Fcal(m)} c_f(q_f); \\
\mathsf{MS}_m(\v{q}, \v{r}) &:= r_m \cdot p_m\left(r_{m} + \sum_{f \in \Fcal(m)} q_{f} \right),
\end{align*}
The quantities $\mathsf{CS}_m$, $\mathsf{PS}_m$, and $\mathsf{MS}_m$ admit natural interpretations. $\mathsf{CS}_m$ equals the consumer surplus in market $m$. $\mathsf{PS}_m$ equals the collective producer surplus of the firms supplying in that market. Finally, $\mathsf{MS}_m$ equals the merchandizing surplus of the market maker from supplying in that market.

The parameterized class of market maker payoff functions defined in \eqref{eq:defPi} encompasses a wide class of common objectives. To illustrate a few, consider the following definitions.
\begin{align}
\v{\theta}^{\mathsf{SW}} := \left( 1, 1, 1 \right), \quad \v{\theta}^{\mathsf{CS}} := \left( 1, 0, 0 \right),  \quad \v{\theta}^{\mathsf{RSW}} := \left( 1, 0, 1 \right), \quad \v{\theta}^{\mathsf{MS}} := \left( 0, 0, 1\right).\label{eq:thetaChoice}
\end{align}
The payoff function with $\v{\theta}^{\mathsf{SW}}$ as the design parameter is the Walrasian social welfare that is widely used in many centrally managed networked marketplaces, including wholesale electricity markets. In the same vein, $\Pi\left(\v{q}, \v{r}, \v{\theta}^{\mathsf{CS}}\right)$ is the collective consumer surplus across all markets, and hence, defines a pro-consumer design choice by the market maker. Another common choice is $\Pi\left(\v{q}, \v{r}, \v{\theta}^{\mathsf{RSW}}\right)$, the \emph{residual social welfare}, which equals the social welfare less the collective producer surplus of all firms. By maximizing the residual social welfare, the market maker hopes to strike a balance in optimizing the components of the social welfare that do not accrue to firms. 
In contrast with $\v{\theta}^{\mathsf{SW}}$, $\v{\theta}^{\mathsf{CS}}$, and $\v{\theta}^{\mathsf{RSW}}$, the choice of $\v{\theta}^{\mathsf{MS}}$ corresponds to a profit-maximizing market maker.

Note that we do not model any variable costs associated with transporting the commodity through the network.
However, as long as the variable costs are convex in $\v{r}$, most of our results will continue to hold.

\textbf{A motivating example:} Many networked marketplaces with market makers that govern transport can be described by the model discussed above, but to provide a concrete motivating example for use throughout this paper, we consider the case of \emph{wholesale electricity markets}.  We illustrate our results with this example in Section \ref{sec:approximateDesign}.

\begin{figure}[t]
\centering
\includegraphics[width=0.5\textwidth]{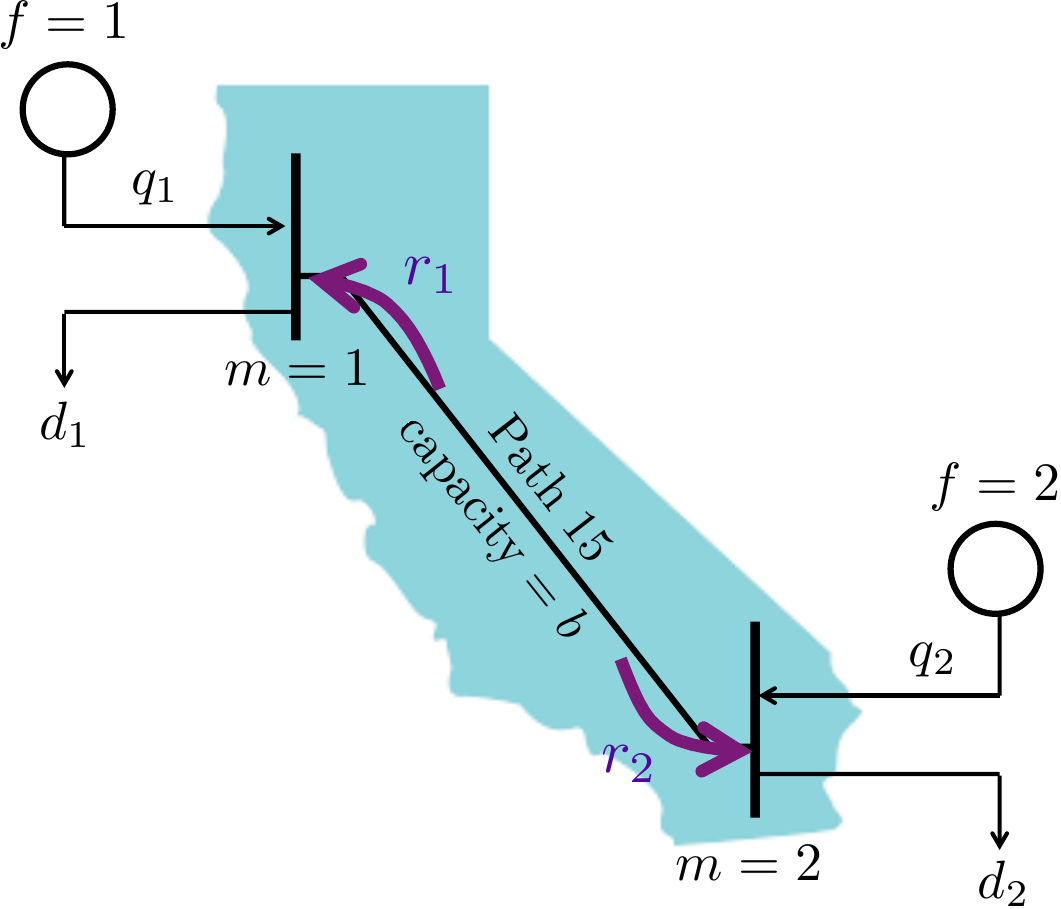}
\caption{Example of a two-market two-firm networked marketplace. This example reperesents a caricature of the wholesale electricity market in California. Here, northern and southern California are represented as two nodes connected by a transmission line - Path 15 - that is often congested (see~\cite{sweeney2008california}).}
\label{fig:2Node}
\end{figure}

Organized wholesale electricity markets in the US are managed by a regulatory entity known as an Independent System Operator (ISO). The role of the ISO is to facilitate efficient exchange of power between supply and demand while ensuring that power flows through the network satisfy the operating constraints of the grid.  Thus, the ISO plays the role of the market maker in our model.

To illustrate the model, consider the two-node network in Figure \ref{fig:2Node}. Here, northern and southern California are modeled as two nodes connected by a transmission line -- Path 15. Assume, for simplicity, that there is one generator at each node and the transmission line has  capacity $b \in \Rset_+$. The California Independent System Operator (CAISO) serves as the market maker, governing transport, and seeks to maximize social welfare through reallocating generation.

We can model the strategic interactions in this simple example as a game where, there are two markets $\Mcal = \{1,2\}$ with inverse linear demand functions $p_1(d_1) = \alpha_1 - \beta_1 d_1$ and $p_2(d_2) = \alpha_2 - \beta_2 d_2$, and two firms $\Fcal (1) = \{1\}$ and $\Fcal (2) = \{2\}$ with cost functions $c_1(q_1)$ and $c_2(q_2)$, respectively.  The set of feasible reallocations by the market maker is $\Pcal = \{ \v{r} \in \Rset^2 :  \lvert r_1 \rvert \leq b, \ \lvert r_2 \rvert \leq b \}$. The market maker's payoff is the social welfare, i.e., the design parameter is $\v{\theta}^{\mathsf{SW}}$.

\textbf{Equilibrium definition:}  We conclude this section by formally describing the networked Cournot competition as the parameterized game $\Gset(\v{\theta})$ among the firms in $\Fcal$ and the market maker. Each firm $f$ plays $q_f \in \Rset_+$ and its payoff is given by $\pi_f$. The market maker plays $\v{r} \in \Pcal$ such that $\bone^\top \v{r} = 0$. Its payoff is given by $\Pi$ that is parameterized by $\v{\theta}$.

We focus our analysis on the Nash equilibria outcomes, which are defined as follows:  $(\v{q}, \v{r}) \in \Rset^{| \Fcal |}_+ \times \Pcal$, satisfying $\bone^\top \v{r} = 0$, comprises a \emph{Nash equilibrium} of $\Gset(\v{\theta})$, if
\begin{align*}
\pi_{f}(\v{q}, \v{r}) & \geq \pi_{f}(q'_f, \v{q}_{-f}, \v{r}), \ \ \text{for all } q'_f \in \Rset_+, \\
\Pi(\v{q}, \v{r}; \v{\theta}) & \geq \Pi(\v{q},\v{r'}; \v{\theta}), \ \ \text{for all } \v{r'} \in \mathcal{P}, \ \bone^\top \v{r'} = 0.
\end{align*}

This work considers a simultaneous move game between the market maker and the firms. Alternate timing choices require the market maker and the firms to engage in a Stackelberg game; see \cite{metzler2003nash} for a discussion. We adopt the simultaneous move game for two reasons. First, it has been extensively studied in the electricity market literature that serves as our motivating example, e.g.,~\cite{jing1999spatial,yao2004computing,YaoWillemsOrenAdler2005,neuhoff2005network,barquin08cournot}. Such a model has been known to explain price behaviors observed in practice \cite{willems2009cournot}. Second, the focus of this paper is on studying the effect of $\v{\theta}$ on the Nash equilibria of $\Gset(\v{\theta})$ that represents the impact of market maker design. Characterization of the equilibria itself (with a social planner as the market maker) can be considerably challenging in Stackelberg models, as revealed in \cite{xuefficiency}.


\newcommand{\Gt}{{\Gset(\v{\theta})}}
\newcommand{\Ct}{{\Cset(\v{\theta})}}
\newcommand{\vt}{{\v{\theta}}}

\section{Characterizing the Nash Equilibria}
\label{sec:NEchar}

In this section, we describe our first set of results, which provide characterizations of the equilibria outcomes, and contrast the equilibrium in our networked Cournot marketplace to non-networked Cournot models.  Then, in Section \ref{sec:design}, we use the characterizations provided here to inform the design of the market maker.

\subsection{Existence and uniqueness}

Classical Cournot competition among a set of firms in a single market with inverse linear demand functions is known to be a potential game (see \cite{slade1994does} and \cite{monderer1996potential}) and, recently, this property has been shown to extend to a form of networked Cournot competition, as shown in \cite{abolhassani2014network}).  These characterizations are powerful, since they allow results about equilibrium existence and uniqueness to be derived through analysis of the underlying potential function of the game.  However, the results in \cite{abolhassani2014network} focus on a form of networked competition over bipartite graphs with no market maker; thus they do not apply to the model we consider here.  But, given the results for these classical and networked Cournot models, an optimistic reader expect a similar conclusion for the model we consider.  In the results that follow, we show that this is true in some situations -- under some assumptions, we show that the model we consider yields a weighted potential game -- however, the structure of the game is more complex in general.

Before stating our results, we formally define a weighted potential game. Consider an $N$-player game with Euclidean strategy sets $\Scal_1, \ldots, \Scal_N$ and payoffs $\varphi_i : \Scal \to \Rset$ for each $i=1, \ldots N$. Define $\Scal := \times_{i=1}^N \Scal_i$. It is said to be a \emph{weighted potential game}, if there exists a vector of weights $\v{w} \in \Rset^N_{++}$ and a potential function $\Phi : \Scal \to \Rset$ that satisfies
\begin{align*}
\Phi(\v{x}_i,  \v{x}_{-i}) - \Phi(\v{x}'_i,  \v{x}_{-i}) = w_i \cdot \left[\varphi_i(\v{x}_i,  \v{x}_{-i}) - \varphi_i(\v{x}'_i, \v{x}_{-i})\right]
\end{align*}
for each $(\v{x}_i, \v{x}_{-i}) \in \Scal$ and $(\v{x}_i^\prime, \v{x}_{-i}) \in \Scal$, and $i=1,\ldots,N$.

Our first result highlights that, for some design parameters $\v{\theta}$, our networked competition model is a potential game whose potential function is a perturbed version of the market maker's payoff.

\begin{theorem}
\label{thm:pot}
If $\theta_M+\theta_P-\theta_C > 0$,
then $\Gt$ is a weighted potential game with the potential function
$\hat{\Pi}(\v{q}, \v{r}; \v{\theta})$, given by
\begin{align}
\hat{\Pi}(\v{q}, \v{r}; \v{\theta})
&:= {\Pi}(\v{q}, \v{r}; \v{\theta})  - (\theta_M - \theta_P) \sum_{m\in\Mcal} \frac{\beta_m}{2} \left( \sum_{f \in \Fcal(m)} q_f \right)^2 \notag\\
& \qquad -\sum_{f \in \Fcal} \left[  \left( \theta_M + \theta_P - \theta_C \right) \frac{\beta_{\Mcal(f)}}{2} q_f^2 +  (\theta_
C - \theta_M) \left( \alpha_{\Mcal(f)} q_f - c_f (q_f)  \right)\right].
\label{eq:defPiHat}
\end{align}
\end{theorem}

A proof of this result is provided in Appendix \ref{sec:proof.pot}. The fact that $\Gt$ is a potential game highlights that it has a number of favorable properties. In particular, the optimizers of the following problem can be used to infer existence -- and in some cases, uniqueness -- of equilibria of $\Gt$.
\begin{align}
\Cset(\v{\theta}) \ : \ & \underset{\v{q}, \v{r}}{\text{maximize}} \   \hat{\Pi}(\v{q}, \v{r}; \v{\theta}), \ \  \text{subject to}   \quad
\v{q} \in \Rset^{| \Fcal |}_+, \v{r} \in \Pcal, \ \bone^\top \v{r} = 0.
\label{eq:Ctheta}
\end{align}
In addition, the above problem can be solved efficiently to compute an equilibrium for a wide variety of cost functions for the firms (e.g., increasing linear or convex quadratic costs). Finally, many natural learning dynamics are known to converge to an equilibrium in potential games.  See \cite{monderer1996potential} and more recent publications, e.g., \cite{fudenberg1998learning, young2004strategic, shamma2004unified, marden2009cooperative, marden2009payoff} for a  discussion on the topic.

The possible characterization of existence of equilibria from Theorem \ref{thm:pot} is not complete.  It turns out that, for many design parameters, the structure of the game is more complex and, in particular, the game is not a potential game.  Despite this, in such cases a Nash equilibrium may still be guaranteed to exist.  The theorem below provides a more complete view of existence and uniqueness of equilibria.

\begin{theorem}
\label{thm:main}
Suppose $\Pcal$ is compact and convex, and let
\begin{align}
\gamma := 1 - \min_{m \in \Mcal}\left( 1 + \sum_{f \in \Fcal(m)} \frac{\beta_m}{\beta_m + \inf_{q_f\geq 0} c_f''(q_f)} \right)^{-1}.
\label{eq:uniCond}
\end{align}
\begin{enumerate}[(a)]
\item If $2\theta_M - \theta_C \geq 0$ or $\theta_M+\theta_P-\theta_C > 0$, then $\Gt$ has a Nash equilibrium.
\item If $2\theta_M - \theta_C \geq \gamma \cdot \left(\theta_M+\theta_P-\theta_C\right) > 0$, then the set of Nash equilibria of $\Gt$ is nonempty, and is identical to the set of optimizers of $\Ct$. Furthermore, if the inequalities are strict, then $\Gt$ has a unique Nash equilibrium.
\end{enumerate}
\end{theorem}
The formal proof is deferred till Appendices \ref{sec:proof.main.a} and \ref{sec:proof.main.b}. Existence guarantee for design parameters with $\theta_M+\theta_P-\theta_C > 0$ utilizes the potential game characterization from Theorem \ref{thm:pot}. For the other condition on existence, we appeal to a classical result due to Rosen in \cite{rosen1965existence}. It requires the market maker's payoff to be concave in $\v{r}$, a property that holds when $2\theta_M - \theta_C \geq 0$. This result highlights that $\Gt$ has additional structure for design parameters beyond where it is a potential game. Within the assumptions of part (b), we demonstrate that the potential function is concave (and strongly concave when the inequality describing the condition is strict). Our proof utilizes the fact that all equilibria of a potential game with a concave differentiable potential function are optimizers of the said potential. We prove the last statement in the appendix, closely following Neyman's arguments in \cite{neyman1997correlated}. It allows us to infer uniqueness of equilibria for part of the design space.

Theorem \ref{thm:main}(b) is derived for a constrained network with compact $\Pcal$. Our proof technique applies more generally to unconstrained networks as well. We record this result in the following corollary.
\begin{corollary}
\label{corr:unboundedP}
Suppose $\Pcal$ is closed and convex. If $2\theta_M - \theta_C > \gamma \cdot \left(\theta_M+\theta_P-\theta_C\right) > 0$, then $\Gt$ has a unique Nash equilibrium, given by the unique optimizer of $\Ct$, where $\gamma$ is defined in \eqref{eq:uniCond}.
\end{corollary}

In Figure \ref{fig:simplex}, we visualize the regions defined by the conditions in Theorem \ref{thm:main}.  Notice that the equilibria of $\Gset(\v{\theta})$ is invariant under a positive scaling of $\vt$. Thus, we restrict our attention to $\vt$ varying over the 3-dimensional simplex $\Delta := \{ \v{\theta} \in \Rset^3_+ \ : \ \theta_C + \theta_P + \theta_M = 1 \}$. The conditions required on $\vt$ in Theorem \ref{thm:main}(b) depend on $\gamma$, that in turn depends on the nodal market demand functions and the firms' cost functions. Costs being convex, $c_f''$ is nonnegative, and hence, we get
$$ \gamma \leq  \max_{m \in \Mcal} \frac{| \Fcal(m) |}{1 + | \Fcal(m) |} < 1.$$
Note that $\gamma = \frac{1}{2}$ when each market has only one firm and costs are increasing linear functions. For illustrative purposes, we choose $\gamma=\frac{1}{2}$ to portray the various regions of $\Delta$ in Figure \ref{fig:simplex}, where $\Gt$ has different properties.

\tikzset{font = \small}
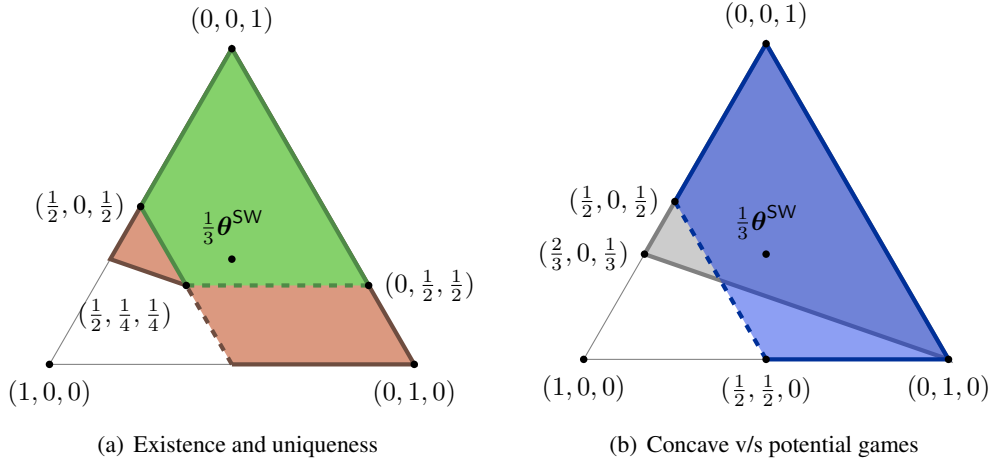
\begin{figure}[t]
    \centering
	\subfigure[Existence and uniqueness]{
                {{ \begin{tikzpicture}
                [scale = 2.8, thick,
                      acteur/.style={
                        circle,
                        fill=black,
                        thick,
                        inner sep=1pt,
                        minimum size=0pt
                      }]

                \definecolor{exist}{rgb}{0.86, 0.60, 0.50}
                \definecolor{existB}{rgb}{0.43, 0.30, 0.25}
                \definecolor{unique}{rgb}{0.52, 0.82, 0.42}
                \definecolor{uniqueB}{rgb}{0.31, 0.54, 0.28}

                \coordinate (1) at (0, 0);
                \coordinate (2) at (1.73205, 0);
                \coordinate (3) at (0.866025, 1.5);
                \coordinate (4) at (0.866025, 0.5);
                \coordinate (5) at (0.866025, 0);
                \coordinate (6) at (0.649519, 0.375);
                \coordinate (7) at (0.288675,0.5);
                \coordinate (8) at (1.51554, 0.375);
                \coordinate (9) at (0.433013,0.75);

                \draw[color=gray, thin] (1) -- (2) -- (3) -- (1);

                \fill[fill=exist, opacity=1]  (7) -- (6) -- (5) -- (2) -- (3) -- cycle;
                \draw[color=existB, ultra thick, dashed] (5) --  (6);
                \draw[color=existB, ultra thick] (5) -- (2) -- (3) -- (7) -- (6);

                \fill[fill=unique, opacity=1]  (6) -- (9) -- (3) -- (8) -- cycle;
                \draw[color=uniqueB, ultra thick, dashed] (8) -- (6);
                \draw[color=uniqueB, ultra thick] (6) -- (9) -- (3) -- (8);

                \node (1) at (0, 0) [acteur,label=below:{$(1, 0, 0)$}]{};
                \node (2) at (1.73205, 0) [acteur,label=below:{$(0, 1, 0)$}]{};
                \node (3) at (0.866025, 1.5) [acteur,label=above:{$(0, 0, 1)$}]{};
                \node (4) at (0.866025, 0.5) [acteur,label=above:{$\frac{1}{3}\v{\theta}^{\sf SW}$}]{};
                \node (6) at (0.649519, 0.375) [acteur,label=south west:{$(\frac{1}{2}, \frac{1}{4}, \frac{1}{4})$}]{};
                \node (8) at (1.51554, 0.375) [acteur,label=right:{$(0,\frac{1}{2}, \frac{1}{2})$}]{};
                \node (9) at (0.433013,0.75) [acteur,label=left:{$(\frac{1}{2}, 0, \frac{1}{2})$}]{};
                \end{tikzpicture}
                }}\label{fig:ExUn}}
    \ \
    \subfigure[Concave v/s potential games]{
                {{ \begin{tikzpicture}
                [scale = 2.8, thick,
                      acteur/.style={
                        circle,
                        fill=black,
                        thick,
                        inner sep=1pt,
                        minimum size=1pt
                      }]
                \definecolor{rosen}{rgb}{.8,.8,.8}
                \definecolor{rosenB}{rgb}{.5,.5,.5}
                \definecolor{potential}{rgb}{0, 0.16, 0.9}
                \definecolor{potentialB}{rgb}{0, 0.19, 0.6}

                \coordinate (1) at (0, 0);
                \coordinate (2) at (1.73205, 0);
                \coordinate (3) at (0.866025, 1.5);
                \coordinate (4) at (0.866025, 0.5);
                \coordinate (5) at (0.866025, 0);
                \coordinate (6) at (0.649519, 0.375);
                \coordinate (7) at (0.288675,0.5);
                \coordinate (8) at (1.51554, 0.375);
                \coordinate (9) at (0.433013,0.75);

                \draw[color=gray, thin] (1) -- (2) -- (3) -- (1);

                \fill[fill=rosen]  (7) -- (2) -- (3) -- cycle;
                \draw[color=rosenB, ultra thick]  (7) -- (2) -- (3) -- cycle;

                \fill[fill=potential, opacity=0.45]  (5) -- (2) -- (3) -- (9) -- cycle;
                \draw[color=potentialB, ultra thick, dashed] (9) --  (5);
                \draw[color=potentialB, ultra thick] (5) --  (2) -- (3) -- (9);

                \node (1) at (0, 0) [acteur,label=below:{$(1, 0, 0)$}]{};
                \node (2) at (1.73205, 0) [acteur,label=below:{$(0, 1, 0)$}]{};
                \node (3) at (0.866025, 1.5) [acteur,label=above:{$(0, 0, 1)$}]{};
                \node (4) at (0.866025, 0.5) [acteur,label=above:{$\frac{1}{3}\v{\theta}^{\sf SW}$}]{};
                \node (5) at (0.866025, 0) [acteur,label=below:{$(\frac{1}{2}, \frac{1}{2}, 0)$}]{};
                \node (7) at (0.288675,0.5) [acteur,label=left:{$(\frac{2}{3}, 0, \frac{1}{3})$}]{};
                \node (9) at (0.433013,0.75) [acteur,label=left:{$(\frac{1}{2}, 0, \frac{1}{2})$}]{};

                \end{tikzpicture}
                }}
                \label{fig:rosenPot}}
    		\caption{(a) An illustration of Theorem~\ref{thm:main} for $\vt \in \Delta$. A Nash equilibrium may not exist in the unshaded region, it exists but may not be unique in the brown region, and it is unique and is given by the unique optimizer of $\Ct$ in the green region. (b) An illustration of Theorem~\ref{thm:main}(a) for $\vt \in \Delta$. The grey region is defined by $2\theta_M - \theta_C \geq 0$, where a Nash equilibrium exists owing to a variant of $\Gt$ being a concave game. The blue region is defined by $\theta_M + \theta_P - \theta_C > 0$, where a Nash equilibrium exists because $\Gt$ is a potential game. Dotted line segments on the boundaries of various sets do not belong to the respective sets.}
    \label{fig:simplex}
\end{figure}



Theorems \ref{thm:pot} and \ref{thm:main} provide sufficient conditions for equilibrium existence and uniqueness, but do not address the question of necessity or tightness. To provide some insight into necessity, we provide examples in Appendix \ref{sec:two-node} to highlight that each of the properties may fail to hold if the respective conditions are not met.

\subsection{Example with linear costs and homogeneous demands}
\label{sec:Pfull}

To this point our results have focused only on existence and uniqueness.  We now provide a more detailed characterization of the equilibria.  Specifically, our goal is to study how the equilibria varies with $\vt$.


Without making stronger assumptions on the nature of the game, such a characterization is difficult.  To allow interprebility of the results, we focus on a restricted setting where each market has a single firm with linear increasing cost and the markets have identical linear demand functions. Additionally, we focus on the case of an unconstrained network, i.e., $\Pcal = \Rset^{| \Mcal |}$. It is possible to provide a more general characterization at the expense of interprebility.

In this setting, we are able to offer explicit formulae for the unique Nash equilibrium of $\Gt$ under a subset of the design parameters in Proposition \ref{prop:Pfull}. Importantly, this characterization allows us to contrast the result of competition in the networked marketplace we consider with two cases of particular interest: (a) competition in a collection of non-networked markets, i.e., a setting without transport between markets, and (b) competition in an aggregated market, i.e., a setting where the markets are merged into a single aggregate marketplace without a market maker. The comparison with (a) provides insight into the efficiency of the network  and the comparison with (b) provides insight into the efficiency of the market maker.

Consider $\Gt$ on an unconstrained network ($\Pcal = \Rset^{| \Mcal |}$) joining a collection of markets. Each market has a single firm that supplies in it, and has linear costs $c_f(q_f) := C_f q_f$ with $C_f > 0$. The markets have spatially homogeneous inverse linear demand functions, given by $p_m(d_m)=\alpha - \beta d_m$, for each $m\in\Mcal$, where $\alpha,\beta > 0$. Denote by $\v{C}$, the vector of marginal costs. Its mean and standard deviation are given by
\begin{align}
\bar{C} := \frac{1}{|\Fcal|}\sum_{f\in\Fcal}C_f, \,\, \text{ and }\,\,  \sigma_C := \sqrt{\frac{1}{|\Fcal|}\sum_{f\in\Fcal}(C_f - \bar{C})^2},\label{eq:Cpars}
\end{align}
respectively. We have the following result on this parameterized family of games $\Gset^u(\vt; \v{C}, \alpha, \beta)$.

\begin{proposition}
\label{prop:Pfull}
Consider $\Gset^u(\vt; \v{C}, \alpha, \beta)$, where $\bar{C}$ and $\sigma_C$ are as defined in \eqref{eq:Cpars}. If $2\theta_M-\theta_C >\frac{1}{2} \left(\theta_M + \theta_P - \theta_C\right) > 0$ and $\alpha \geq \left(1 + \kappa(\vt)\right) \max_{f \in \Fcal} C_{f} - \kappa(\vt)\bar{C}$, then $\Gset^u(\vt; \v{C}, \alpha, \beta)$ has a unique Nash equilibrium, given by
\begin{align}
q_{f} &= \frac{1}{2\beta}\left[\alpha - \bar{C} - \left(1 + \kappa(\vt)\right) \left(C_{f} - \bar{C}\right)\right],
\label{eq:Pfull.q}
\\
r_{\Mcal(f)} &= \frac{\kappa(\v{\theta})}{\beta}\left(C_{f} - \bar{C}\right),
\label{eq:Pfull.r}
\end{align}
for each $f \in \Fcal$, where $\kappa(\v{\theta}) := \frac{\theta_M + \theta_P - \theta_C}{3 \theta_M - \theta_P - \theta_C}$. Moreover, the social welfare at the unique equilibrium is
\begin{align*}
\sum_{m\in\Mcal}\left[\mathsf{CS}_m(\v{q},\v{r}) + \mathsf{PS}_m(\v{q},\v{r}) + \mathsf{MS}_m(\v{q},\v{r})\right]
&=
\frac{3|\Fcal|}{8\beta}\left[(\alpha-\bar{C})^2 + \sigma_C^2 + \frac{1}{3}\kappa(\v{\theta})(6 - \kappa(\v{\theta})) \sigma_C^2\right].
\end{align*}
\end{proposition}

%


A proof is given in Appendix \ref{sec:proof.Pfull} that leverages Corollary \ref{corr:unboundedP} together with the Karush-Kuhn-Tucker (KKT) optimality conditions for $\Ct$.
We glean a few insights from the above result. Equations \eqref{eq:Pfull.q} and \eqref{eq:Pfull.r} reveal that the production of a firm $q_f$ and the market-maker's supply in the market served by that firm $r_{\Mcal(f)}$, both depend on the marginal cost of the firm $C_f$ relative to the average marginal cost of all firms $\bar{C}$. Under the conditions of Proposition \ref{prop:Pfull}, one can show that $\kappa(\vt) > 0$. Hence, the firms' productions are in fact ordered inversely by their marginal costs. Moreover, the market maker buys from markets having firms with lower marginal costs and supplies to markets with higher ones.
The total production by all firms, however,  is independent of $\vt$, and is given by
\begin{align}
\label{eq:PfullTotq}
\sum_{f \in \Fcal} q_f
=
\frac{|\Fcal|}{2\beta}\left( \alpha -  \bar{C} \right).
\end{align}
The market maker's design choice only influences the relative production between the firms and the quantities supplied by the market maker to various markets.

Proposition \ref{prop:Pfull} also lets us investigate the efficiency of the equilibrium. We unravel the impact of the design choice on the competitiveness of the market by studying the effect of the design parameter on the social welfare at the unique equilibrium. As we remarked earlier, a popular choice of $\vt$ for a regulated marketplace such as the wholesale electricity markets is $\vt^{\sf SW}$ defined in \eqref{eq:thetaChoice}, i.e., the market maker optimizes the social welfare function. Then, $\kappa\left(\vt^{\sf SW}\right) = 1$. We notice that the social welfare at the unique Nash equilibrium increases with $\kappa(\vt)$ over the interval $[1,3]$. Moreover, it is easy to construct a $\vt$ that satisfies the conditions in Proposition \ref{prop:Pfull} with $1 < \kappa(\vt) < 3$. So, if maximizing the equilibrium welfare is indeed the design goal, $\vt^{\sf SW}$ is \emph{not} the optimal design choice.

One can ask how much efficiency is lost by naively choosing the design parameter $\vt^{\sf SW}$. To address this, we compute the equilibrium welfare with $\vt^{\sf SW}$ and obtain $(\alpha-\bar{C})^2 + \frac{8}{3}\sigma_C^2$. Then, we use the fact that $\kappa(\vt)(6-\kappa(\vt)) \leq 9$ to derive the following upper bound on the ratio of the largest attainable welfare at an equilibrium to that obtained with $\vt^{\sf SW}$.
\begin{align*}
\frac{(\alpha-\bar{C})^2 + \sigma_C^2 + \frac{1}{3}\kappa(\vt)(6-\kappa(\vt))\sigma_C^2}{(\alpha-\bar{C})^2 + \frac{8}{3}\sigma_C^2}
\leq
\frac{1 + 4\left(\frac{\sigma_C}{\alpha-\bar{C}}\right)^2}{1 + \frac{8}{3}\left(\frac{\sigma_C}{\alpha-\bar{C}}\right)^2}  \leq \frac{3}{2}.
\end{align*}
The last step uses the fact that $(1+4x)/\left(1 + \frac{8}{3}x\right)$ increases in $x \geq 0$, and approaches $\frac{3}{2}$ as $x \to \infty$.


When $\vt$ is varied such that $\kappa(\vt)$ increases from one, we have already argued that the  welfare at the equilibrium increases. Who stands to benefit from such an increase? Is it the consumers, the producers, or the market maker? Recall that a metric of consumer benefit is the aggregate consumer surplus $\sum_{m \in \Mcal} \sf{CS}_m(\v{q}, \v{r})$ at equilibrium. Similarly, the aggregate producer surplus and the merchandising surplus at equilibrium measure the benefits to the producers and the market maker, respectively. One can show that the consumer and producer surpluses both increase, when $\vt$ is changed to increase $\kappa(\vt)$ from one. However, the merchandising surplus decreases. Thus, in the framework considered, a design choice that improves the efficiency of the market, does so to the benefit of the consumers and the producers, but at the expense of the market maker.


\subsubsection*{Comparison with non-networked Cournot.}

To study the role of the network, we next analyze the same setting as with the unconstrained network, but with the network removed, i.e., $\Pcal=\{0\}$. Each firm then effectively competes as a monopoly in its own market, and the market maker plays no role.
Call this non-networked Cournot competition as $\Gset^n(\v{C}, \alpha, \beta)$. We inherit the notation $\v{C}, \bar{C}, \sigma_C$ and characterize the equilibria of $\Gset^n(\v{C}, \alpha, \beta)$ in the following result.


\begin{proposition}
\label{prop:nonnetworked}
Consider $\Gset^n(\v{C}, \alpha, \beta)$, where $\bar{C}$ and $\sigma_C$ are as defined in \eqref{eq:Cpars}. If $\alpha \geq \max_{f\in\Fcal} C_f$, then $\Gset^n(\v{C}, \alpha, \beta)$ has a unique Nash equilibrium, given by
$$
q_f^{n} = \frac{1}{2\beta}\left(\alpha - C_f\right).
$$
Moreover, the social welfare at the unique equilibrium is
\begin{align*}
\sum_{f\in\Fcal}\left[ \int_0^{q_f^{n}} p_{\Mcal(f)}(w_f) \, dw_f - C_f q_f^{n} \right]
&=
\frac{3|\Fcal|}{8\beta}\left[\left(\alpha - \bar{C}\right)^2 + \sigma_C^2\right].
\end{align*}
\end{proposition}

The proof is straightforward and is omitted. To compare Propositions \ref{prop:Pfull} and \ref{prop:nonnetworked}, assume that $\alpha$ satisfies the conditions required in both.

Like in the networked marketplace, the production quantities of the firms are ordered inversely by their marginal costs. Also, the total production of the firms at the Nash equilibrium is given by
\begin{align*}
\sum_{f\in\Fcal} q_f^{n} = \frac{|\Fcal|}{2\beta}\left(\alpha - \bar{C}\right),
\end{align*}
which, due to \eqref{eq:PfullTotq}, happens to be identical to that in the networked marketplace.  Thus, the network does not impact the total production of the firms. Instead, the value of the network is reflected in the social welfare at the  equilibrium. 

Taken together, Propositions \ref{prop:Pfull} and \ref{prop:nonnetworked} imply that the equilibrium welfare is \emph{higher} for the networked marketplace for any design choice $\vt$. This aligns with the intuition that a network available for trade improves the efficiency of the marketplace. In the networked setting, recall that the equilibrium welfare is given by $\frac{3|\Fcal|}{8\beta}\left[(\alpha-\bar{C})^2 + \sigma_C^2 + \frac{1}{3}\kappa(\vt)(6-\kappa(\vt))\sigma_C^2\right]$. Again, leveraging the fact that $\kappa(\vt)(6-\kappa(\vt)) \leq 9$, we obtain the following bound on the ratio of the social welfares in the networked and the non-networked case.
\begin{align*}
\frac{(\alpha-\bar{C})^2 + \sigma_C^2 + \frac{1}{3}\kappa(\vt)(6-\kappa(\vt))\sigma_C^2}{(\alpha-\bar{C})^2 + \sigma_C^2}
\leq
\frac{1 + 4\left(\frac{\sigma_C}{\alpha-\bar{C}}\right)^2}{1 + \left(\frac{\sigma_C}{\alpha-\bar{C}}\right)^2} \leq 4,
\end{align*}
since $(1+4x)/\left(1 + x\right)$ increases in $x \geq 0$ and approaches 4 as ${x \to \infty}$. The ratio increases with $\sigma_C$, revealing that the network improves the efficiency more when the firms differ widely in their marginal costs, and nodal markets are able to take advantage of non-local less expensive firms.

\subsubsection*{Comparison with aggregated Cournot.} To study the efficiency of the market maker, we next analyze the same setting, but where the firms are aggregated into a single Cournot market.  This comparison is motivated by the fact that one may hope an efficient market maker can facilitate trade in order to allow the networked marketplace to behave like a single market -- especially when the network is unconstrained.  

Recall that in our example, we considered $| \Mcal |$ markets with identical inverse linear demand functions $p_m(d_m) = \alpha - \beta d_m$ for each $m \in \Mcal$. Then, an aggregation of these markets with a collective demand $d$ admits an inverse linear demand function $p(d) = \alpha - \frac{\beta}{|\Fcal |} d$. Denote the aggregated Cournot competition by $\Gset^a(\v{C}, \alpha, \beta)$, for which we present the following result.



\begin{proposition}
\label{prop:classic}
Consider $\Gset^a(\v{C}, \alpha, \beta)$, where $\bar{C}$ and $\sigma_C$ are as defined in \eqref{eq:Cpars}. If $\alpha \geq \left(1 +  |\Fcal | \right) \max_{f \in \Fcal} C_{f} - |\Fcal| \bar{C}$, then $\Gset^a(\v{C}, \alpha, \beta)$ has a unique Nash equilibrium, given by
\begin{align*}
q_f^{a} = \frac{|\Fcal |}{(1 + |\Fcal|) \beta}\left[ \alpha - \bar{C} - \left(1 +  |\Fcal | \right) \left(C_f - \bar{C}\right) \right].
\end{align*}
Moreover, the social welfare at the unique equilibrium is
\begin{align*}
\int_{0}^{\sum_{f\in\Fcal}q_f^{a}} p(w)\, dw - \sum_{f\in\Fcal} C_f q_f^{a}
&=
\frac{|\Fcal|^2 (2 + |\Fcal|)}{2 (1 + |\Fcal|)^2 \beta}\left[ (\alpha - \bar{C})^2 + \frac{2(1 + |\Fcal|)^2}{2 + |\Fcal|} \sigma_C^2\right].
\end{align*}
\end{proposition}

A proof can be found in \cite{ledvina2010bertrand}, and is omitted for brevity. When comparing the results obtained in Proposition \ref{prop:classic} to that in \ref{prop:Pfull} or \ref{prop:nonnetworked}, assume that $\alpha$ satisfies the conditions delineated in each.

As in each case before, the firms' productions in the aggregated Cournot competition are ordered inversely by their marginal costs. However, in this case the total quantity produced is different. In particular, we have
\begin{align*}
\sum_{f \in \Fcal} q_f^{a}
=
\frac{|\Fcal|^2}{(1 + |\Fcal|)\beta}\left( \alpha - \bar{C} \right).
\end{align*}
Since $\frac{|\Fcal|^2}{1 + |\Fcal|} \geq \frac{|\Fcal|}{2}$, it follows from \eqref{eq:PfullTotq} that the total production quantity in the aggregated Cournot competition is no less than that in the networked marketplace with an unconstrained network. Further, the inequality is strict when $| \Fcal | \geq 2$.

Given increased production, it is natural to expect that the social welfare will be larger in the aggregated Cournot market as well.  This turns out to be true. To compare the social welfare of the aggregated Cournot to our networked marketplace with an unconstrained network, we use the following facts:
(i) $\frac{|\Fcal|^2 (2 + |\Fcal|)}{2 (1 + |\Fcal|)^2 } \geq \frac{3 |\Fcal|}{8}$ for all $| \Fcal | \geq 1$, (ii) $\frac{2(1 + |\Fcal|)^2}{2 + |\Fcal|} \geq 4$ for all $| \Fcal | \geq 2$, and (iii)
$| \Fcal | = 1$ implies $\sigma_C = 0$. These observations, together with $\kappa(\vt) \left( 6 - \kappa(\vt) \right) \leq 9$, yield
\begin{align*}
\frac{|\Fcal|^2 (2 + |\Fcal|)}{2\beta (1 + |\Fcal|)^2 }\left[ (\alpha - \bar{C})^2 + \frac{2(1 + |\Fcal|)^2}{2 + |\Fcal|} \sigma_C^2\right]
&\geq
\frac{3|\Fcal|}{8\beta}\left[(\alpha-\bar{C})^2 + \sigma_C^2 + \frac{\kappa(\v{\theta})(6 - \kappa(\v{\theta}))}{3} \sigma_C^2\right].
\end{align*}
As a result, the social welfare in the aggregate Cournot model is no less than that in the networked Cournot model for all possible choices of the design parameter. The inequality is strict when $| \Fcal | \geq 2$.
Also, $\frac{2(1 + |\Fcal|)^2}{2 + |\Fcal|} \to \infty$ as $| \Fcal | \to \infty$.  Thus, the ratio of equilibrium social welfares in the aggregated market and the unconstrained networked marketplace (with any choice of $\vt$) grows without bound as the number of firms increases. In a sense, the higher the number of firms, the larger the need for transport, leading to a higher efficiency loss due to the market maker's transport.

Therefore, when there are no network constraints, it is more efficient to implement a spatially uniform pricing mechanism based on aggregating demands and supplies across nodes instead of a local pricing mechanism based separately on each node's local demand and supply. Although the equilibrium prices in the latter design may be spatially uniform, firms are able to exploit local market power, and therefore the outcome is less competitive than in an aggregated mechanism.

Our results also imply that limited design instruments may lead to significant inefficiency in markets. The unbounded efficiency loss in networked Cournot competition cannot be attributed completely to firms' strategic behavior since the efficiency loss in aggregated Cournot competition diminishes to zero as $|\Fcal| \to \infty$. Furthermore, the upper bound of $\frac{3}{2}$ on the largest gain in social welfare that can be obtained by optimizing $\vt$ implies that there is potential for significant efficiency gains from considering more sophisticated market maker payoff functions or pricing mechanisms.

The following result collects our findings on the efficiency of equilibria for the case of linear costs and homogeneous demands.

\begin{corollary}
    Suppose $\alpha$ satisfies the conditions in Propositions~\ref{prop:Pfull}-\ref{prop:classic}. Then, the ratio of welfares of $\Gset^u(\vt; \v{C}, \alpha, \beta)$ and $\Gset^u(\vt^{\mathsf{SW}}; \v{C}, \alpha, \beta)$ is bounded above by 3/2. However, the ratio of welfares of $\Gset^a(\v{C}, \alpha, \beta)$ and $\Gset^u(\vt; \v{C}, \alpha, \beta)$ can be unbounded. Moreover, the ratio of welfares of $\Gset^u(\vt; \v{C}, \alpha, \beta)$ and $\Gset^n(\v{C}, \alpha, \beta)$ is bounded above by 4.
\end{corollary}





\section{Market Maker Design}
\label{sec:design}

We build on the characterization results from the previous section to approach the question of market maker design, i.e., to engineer the `right' design parameter $\vt$, when the market maker has a certain design objective. The example considered in Section \ref{sec:Pfull} highlights the importance of this task -- even in simple settings, using $\vt^{\sf SW}$ yields suboptimal outcomes when the goal is to optimize social welfare.

Concretely, the contribution of this section is to find an approximately optimal design parameter, and leverage a \emph{sum of squares} (SOS) framework to bound the suboptimality of that choice. We illustrate our approach on the two-market two-firm example in Figure \ref{fig:2Node}.


\subsection{Formulating the market maker design problem}

Assume that a polynomial $g: \Rset^{|\Fcal |} \times \Rset^{| \Mcal |} \to \Rset$ describes the design objective of the market maker. That is, it would ideally maximize $g(\v{q}, \v{r})$ over the joint strategy set described by $\v{q} \in \Rset^{| \Fcal |}_+, \v{r} \in \Pcal, \bone^\top \v{r} = 0$, if the market maker owned and operated the firms. The goal of market maker design is to drive the outcome of the game towards the same, when playing with a collection of strategic firms. Within the scope of our model, it amounts to finding a design parameter $\vt$ that maximizes $g$ at the Nash equilibrium of $\Gt$. When there are multiple equilibria, we seek to maximize the \emph{worst case} $g$ over all equilibria.

Towards the design goal, we restrict the design space for $\vt$ to the 3-dimensional simplex $\Delta$ without loss of optimality. Optimizing over $\Delta$ is still challenging; the difficulty arises from having to describe the Nash equilibria of $\Gt$ for a candidate $\vt$, if and when they exist. For example, if $\Gt$ has multiple isolated equilibria, optimizing the design choice will become a combinatorial problem. Even if $\Gt$ has a unique equilibrium and the market maker's payoff function is not concave in its own action, then its optimal strategy cannot be described by first-order optimality conditions. Even if that function is concave, computing a Nash equilibrium of $\Gt$ -- and hence, computing $g$ for any candidate $\vt$ -- can be difficult in general.

In light of these challenges, we restrict the search space for $\vt$ to the closed set $\Theta_\ve$, described by
\begin{align*}
\hspace{-6pt}\theta_C, \theta_P, \theta_M \geq 0,  \ \ \theta_C + \theta_P + \theta_M = 1,  \ \ 2\theta_M - \theta_C \geq \ve + \gamma \cdot (\theta_M + \theta_P - \theta_C) \geq (1+\gamma)\cdot \ve,
\end{align*}
for a small $\epsilon > 0$. Theorem \ref{thm:main}(b) implies that $\Gt$ has a unique Nash equilibrium for each $\vt \in \Theta_\ve$ that also equals the unique optimizer of the convex program $\Ct$ in \eqref{eq:Ctheta}.
Hence, the unique equilibrium is exactly characterized by the Karush-Kuhn-Tucker (KKT) optimality conditions for $\Ct$. Let $\v{\lambda}$ denote the collection of Lagrange multipliers associated with $\Ct$. We arrive at the following conceptual market maker design problem over $\Theta_\ve$.
\begin{equation}
\begin{alignedat}{8}& \underset{\v{q}, \v{r}, \v{{\lambda}}, \vt}{\text{maximize}}
	 & &  g(\v{q}, \v{r}), \\
 & \text{subject to}   \qquad
	&& (\v{q}, \v{r}, \v{\lambda}) \text{ satisfies the KKT conditions for } \Ct, \\
	&&& \vt \in \Theta_\ve.
 \end{alignedat} \label{eq:findTheta.2}
\end{equation}
%
Identify an optimizer $\vt^*$ of \eqref{eq:findTheta.2} as an optimal design choice.
As we illustrated in Section \ref{sec:Pfull} through an example, even if $g(\v{q}, \v{r})=\Pi(\v{q}, \v{r}; \vt^0)$ for some $\vt^0 \in \Theta_\ve$, the design choice $\vt^0$ may not be optimal. 


\subsection{Approximately solving the market maker design problem}
\label{sec:approximateDesign}
The market maker design problem in \eqref{eq:findTheta.2} is a so-called \emph{mathematical program with equilibrium constraints} (MPEC). Such problems are nonconvex and generally hard to solve; see \cite{LuoPangRalph1996,Pieper2001,OutrataKocvaraZowe2013}.
Instead of using existing heuristics to solve MPECs (that often do not come with optimality guarantees), we provide a scheme to find an approximate solution of \eqref{eq:findTheta.2} and bound the resulting optimality gap.

Assume henceforth that the cost functions $c_f, f\in\Fcal$ are polynomials for which $\Ct$ can be solved efficiently to yield the equilibrium. For example, when these costs are quadratic, $\Ct$ can be solved as a convex quadratic program.
Any metaheuristic can be used to explore the space $\Theta_\ve$ to locate where $g$ is maximized at the equilibrium, e.g., grid search, simulated annealing, Monte-Carlo sampling, etc. We adopt a grid search and maximize $g$ over a finite uniform discretization of $\Theta_\ve$ to obtain $\vt_{\max}$. In general, $\vt_{\max}$ will be a suboptimal design choice. The difference between $g$ evaluated at the Nash equilibrium of $\Gcal(\vt_{\max})$ and that of $\Gcal(\vt^*)$ measures the optimality gap of $\vt_{\max}$. In what follows, we present a way to compute a bound on this gap. More precisely, we provide a hierarchy of successively tighter upper bounds  on the optimal cost of \eqref{eq:findTheta.2}. At any level of the hierarchy, such an upper bound can be efficiently computed, and it yields a bound on the optimality gap of $\vt_{\max}$.

To define the hierarchy, we show that \eqref{eq:findTheta.2} can be written as a \emph{polynomial optimization}\footnote{A polynomial optimization problem seeks to optimize a polynomial function of its arguments over a feasible set described by polynomial equalities and inequalities.} problem in $\v{z} := \left( \v{q}, \v{r}, \v{{\lambda}}, \vt \right)$ over a compact feasible set, if $\Pcal$ is a polytope, i.e.,
$$ \eqref{eq:findTheta.2} \ \ \equiv \ \  \text{maximize} \ g(\v{z}), \text{ subject to } h_i(\v{z}) \geq 0, \quad i = 1,\ldots, I,$$
where $h_1, \ldots, h_{I}$ are polynomials, and $h_I(\v{z}) = {Z} - \| \v{z} \|_2$ for some ${Z} > 0$. We have slightly abused notation in writing $g(\v{z})$ to denote $g(\v{q}, \v{r})$.
See Appendix \ref{sec:proof.SOS} for its proof.\footnote{Our proof technique is tailored for $\Pcal$ being a polytope. For a general compact convex semi-algebraic set $\Pcal$, one can utilize the framework in \cite{jeyakumar2015convergent} that formulates problems such as \eqref{eq:findTheta.2} using Fritz-John optimality conditions in place of the Karush-Kuhn-Tucker conditions for $\Ct$.}
%
%
Now, employ the so-called \emph{Lasserre hierarchy} (see \cite[Chapter 4]{lasserre2009moments}) to provide upper bounds on the optimal cost of \eqref{eq:findTheta.2} as follows.
Call a polynomial as a \emph{sum of squares} (SOS), if it can be expressed as a sum of other squared polynomials, and define
\begin{equation}
\begin{alignedat}{8} v_d^* := \  &  \underset{t, \sigma_0, \ldots, \sigma_I}{\text{minimize}}
	 & &  t, \\
 & \text{subject to}   \quad
	&& t = g + \sigma_0 + \sigma_1 h_1 + \ldots, \sigma_I h_I,\\
	&&& \deg (\sigma_0) \leq 2 d, \ \deg (\sigma_i h_i) \leq 2 d, \ i=1,\ldots,I,\\
	&&& t \in \Rset, \ \sigma_0, \ldots, \sigma_I \text{ are SOS},
 \end{alignedat} \label{eq:findTheta.4}
\end{equation}
for a positive integer $d \geq \frac{1}{2} \max_{i=1,\ldots,I} \deg(h_i)$. Here, $\deg(\cdot)$ stands for the degree of the respective polynomial. Theorem 4.1 in \cite{lasserre2009moments} guarantees that $v_d^*$ monotonically decreases to the optimal cost of the market maker design problem in \eqref{eq:findTheta.2}. Further, the SOS constraints on the polynomials in \eqref{eq:findTheta.4} can be written as linear matrix inequalities in the coefficients of the same polynomials (see \cite{lasserre2009moments}). As a result, $v_d^*$ can be computed efficiently using a semidefinite program. If $\vt_{\max}$ is an approximately optimal design choice for \eqref{eq:findTheta.2}, then the difference of $v_d^*$ and $g$ evaluated at the equilibrium of $\Gcal(\vt_{\max})$ bounds the optimality gap of $\vt_{\max}$.


To illustrate our scheme for market maker design, consider again the two-market two-firm example discussed in Section \ref{sec:model} with the following parameters.
$$ c_1(q_1) := \frac{1}{2} q_1, \ \ c_2(q_2) := \frac{1}{4} q_2, \ \ p_m(d_m) := 1 - d_m, \text{ for } m =1,2.$$
The line capacity is $b = \frac{1}{2}$. With the social welfare function as the market clearing objective, i.e., $\vt^{0} =  \frac{1}{3} \vt^{\sf SW}$,  Appendix \ref{sec:two-node} yields $q_1 \!=\! \frac{3}{16}, \, q_2 \!=\! \frac{7}{16},\, r_1 \!=\! - r_2 \!=\! \frac{1}{8}$ as the unique Nash equilibrium of $\Gset\left(\vt^{0}\right)$, where welfare equals $\frac{83}{256} \approx 0.324$. A naive grid search over $\Theta_\ve$ with $\ve = 0.001$ yields $\vt_{\max} = (0.027, 0.627, 0.346)^\top$. Social welfare at the unique equilibrium of $\Gset(\vt_{\max})$ is $0.339$, which is higher than $0.324$ obtained at that of $\Gset\left(\vt^{0}\right)$. We further obtain an upper bound $v_1^* = 0.340$ on the maximum attainable welfare at an equilibrium over $\Theta_\ve$.
The upper bound being close to the social welfare at $\vt_{\max}$ suggests that $\vt_{\max}$ is indeed a good design choice. The design, however, falls short of the welfare in a perfectly competitive market, where the optimal allocation is $q_1 \!=\! 0, \, q_2 \!=\! \frac{5}{4},\, r_1 \!=\! - r_2 \!=\! \frac{1}{2}$ with a welfare of $\frac{17}{32} \approx 0.531$. The inability to elicit the competitive outcome stems from the limited design instruments available to the market maker. A market design for the same, however, will typically require completely upending the current mechanism in practice.

The market maker's ability to tailor $\vt$ to its benefit is a form of leader's advantage in Stackelberg competition. Being able to anticipate the market outcomes under different design choices, the market maker can therefore change the rules of the competition to obtain more favorable outcomes for itself. Existing theory on Stackelberg games does not immediately imply that $\vt^{\mathsf{SW}}$ is not the optimal design choice. The optimal design usually depends on the instruments available to the market maker. Our results can be leveraged to both design with limited instruments and also determine the richness of that design space.




\section{Conclusion}

This work considers a Cournot competition of a single commodity among a collection of strategic firms in a centrally managed networked marketplace. The central manager (market maker) facilitates transport of the commodity over an underlying network. The case of wholesale electricity markets is used throughout as our motivating example; our analysis, however, applies more generally to shared economies, supply chains, etc. Of particular interest is understanding the role of the market maker design. That is, we study how the market clearing rule of the market maker influences the Nash equilibrium outcomes of the marketplace.

Our main result (Theorem \ref{thm:main}) characterizes the equilibria outcomes over a parameterized family of market maker designs. We identify the set of design parameters over which an equilibrium is guaranteed to exist and is unique. Then, we exploit our characterization to propose an approach for finding an approximately optimal design choice when the market maker has a specific design objective in mind. A sum of squares based relaxation framework is utilized to bound the optimality gap of our approach. 


We illustrate our results on a two-node network where the market maker maximizes social welfare to clear the market. We demonstrate that maximizing the social welfare as the market clearing rule is not always an optimal design choice when the objective is to maximize social welfare at the outcome of the game. For the example considered, our approximation scheme in fact yields a near optimal design choice.



\section{Proofs}
\label{sec:proofs}

This appendix is dedicated to the formal proofs of the various results.
In the interest of brevity, we use the notation
\begin{align}
\label{eq:defP'}\Pcal' := \Pcal \cap \{ \v{r}:\bone^\top \v{r} = 0\}, \quad \text{and} \quad Q_m = \sum_{f \in \Fcal(m)} q_f
\end{align}
for each $m \in \Mcal$.
Our proofs will leverage two results -- one is standard in the literature and the second one is proven here for completeness. To state these results, we need the following definition.
Consider an $N$-player game with Euclidean strategy sets $\Scal_1, \ldots, \Scal_N$ and payoffs $\varphi_i : \Scal \to \Rset$ for each $i=1, \ldots N$, where $\Scal := \times_{i=1}^N \Scal_i$. A game is said to be \emph{concave}, if $\varphi_i$ is continuous in its arguments and concave in $\v{x}_i$ for each $i=1,\ldots,N$, and $\Scal$ is compact and convex. A game is called a \emph{weighted potential game}, if there exists weights $\v{w} \in \Rset^N_{++}$ and a potential function $\Phi : \Scal \to \Rset$ that satisfies
\begin{align}
\label{eq:pot} \Phi(\v{x}_i,  \v{x}_{-i}) - \Phi(\v{x}'_i,  \v{x}_{-i}) = w_i \cdot \left[\varphi_i(\v{x}_i,  \v{x}_{-i}) - \varphi_i(\v{x}'_i, \v{x}_{-i})\right]
\end{align}
for each $(\v{x}_i, \v{x}_{-i}) \in \Scal$ and $(\v{x}_i^\prime, \v{x}_{-i}) \in \Scal$, and $i=1,\ldots,N$.

\begin{lemma}
\label{lemma:generalGames}
\begin{enumerate}[(a)]
\item A Nash equilibrium always exists in a concave game.
\item For a weighted potential game with a potential function $\Phi$, all maximizers of $\Phi$ over $\Scal$ are Nash equilibria. Moreover, all equilibria are maximizers of $\Phi$ over $\Scal$, if $\Scal$ is convex, and $\Phi$ is continuously differentiable and jointly concave over $\Scal$.
\end{enumerate}
\end{lemma}
\proof{Proof.}
Part(a) is adapted from Rosen's result in \cite[Theorem 1]{rosen1965existence}. For part (b), the definition of $\Phi$ in \eqref{eq:pot} implies that all maximizers of $\Phi$ over $\Scal$ are Nash equilibria (see \cite[Corollary 2.1]{la2016potential}). In the following, we borrow from Neyman's proof in \cite[Theorem 1]{neyman1997correlated} to argue the converse.\footnote{Neyman's result in the corollary to \cite[Theorem 1]{neyman1997correlated} does not directly apply, owing to the requirement that the payoffs need to remain bounded over $\Scal$.}

Assume to the contrary that $\v{x}^* \in \Scal$ is a Nash equilibrium that is \emph{not} a maximizer of $\Phi$ over $\Scal$. Then, there exists $\v{y} \in \Scal$ such that $\Phi(\v{y}) > \Phi(\v{x}^*)$. Appeal to the concavity of the continuously differentiable
\footnote{We call $\Phi$ continuously differentiable on $\Scal$ to mean that there exists a continuous map $\nabla \Phi$ defined on $\Scal$ such that the directional derivative of $\Phi$ at $\v{x}$ towards $\v{v}$ equals
$\langle \nabla\Phi(\v{x}), \v{v} \rangle$
whenever $\v{x}$ and $\v{x} + \v{v}$ are in $\Scal$.
}
function
$\Phi$ to infer
\begin{align*}
\langle \nabla\Phi(\v{x}^*), \v{y} - \v{x}^* \rangle = \lim_{\ve \downarrow 0}  \frac{1}{\ve}\left[{\Phi(\v{x}^* + \ve (\v{y} - \v{x}^*)) - \Phi(\v{x}^*)}\right]
\geq \Phi(\v{y}) - \Phi(\v{x}^*) > 0.
\end{align*}
Here $\langle \cdot, \cdot \rangle$ stands for the usual dot product.
Decompose into player-wise components to write $\v{y} - \v{x}^* = \sum_{i=1}^n \v{y}_i - \v{x}^*_i$ in the above inequality and deduce that
$\sum_{i=1}^N \langle \Phi(\v{x}^*), \v{y}_i - \v{x}^*_i \rangle
 > 0.$
At least one among these summands is strictly positive -- say the $i$-th one. The last observation, together with \eqref{eq:pot}, then yields
\begin{align*}
\lim_{\ve \downarrow 0}  \frac{1}{\ve} \left[{\varphi_i(\v{x}^* + \ve (\v{y}_i - \v{x}_i^*)) - \varphi_i(\v{x}^*)}\right]
= \frac{1}{w_i}\lim_{\ve \downarrow 0}  \frac{1}{\ve} \left[{\Phi(\v{x}^* + \ve (\v{y}_i - \v{x}_i^*)) - \Phi(\v{x}^*)}\right]
= \langle \nabla \Phi(\v{x}^*), \v{y}_i - \v{x}^*_i \rangle
> 0.
\end{align*}
Therefore, a small step from $\v{x}^*$ along $\v{y}_i - \v{x}^*_i$ remains in $\Scal$ and leads to an increase in $\varphi_i$. Said differently, player $i$ improves its payoff from a unilateral deviation from $\v{x}^*$. It contradicts that $\v{x}^*$ is an equilibrium, completing the proof.
\Halmos
\endproof
%


\subsection{Proof of Theorem \ref{thm:pot}.}
\label{sec:proof.pot}
From the definition of $\hat{\Pi}$, it follows that $\hat{\Pi}(\v{q}, \v{r}; \v{\theta})- {\Pi}(\v{q}, \v{r}; \v{\theta})$ does not depend on $\v{r}$. Hence, for every $\v{q} \in \Rset^{|\Fcal|}_+$, we have
\begin{align}
\hat{\Pi}(\v{q}, \v{r}; \v{\theta}) - \hat{\Pi}(\v{q}, \v{r}'; \v{\theta}) = {\Pi}(\v{q}, \v{r}; \v{\theta}) - {\Pi}(\v{q}, \v{r}'; \v{\theta})
\label{eq:diff.r}
\end{align}
for each $\v{r},\v{r}'\in\Pcal'$. Recall that $Q_m := \sum_{f \in \Fcal(m)} q_f$. Upon expanding $\hat{\Pi}(\v{q},\v{r};\v{\theta})$, we then obtain\footnote{Setting $\vert\Mcal\vert = 1$, $\v{r} = \mathbf{0}$, and $\vt=(1,1,1)$, we recover the potential function for classical Cournot competition in \cite{monderer1996potential}.}
\begin{align}
\hat{\Pi}(\v{q},\v{r};\v{\theta})
&=
\left(\theta_M + \theta_P - \theta_C\right) \sum_{m\in\Mcal} \left[ (\alpha_m - \beta_m r_m) Q_m - \sum_{f\in\Fcal(m)} c_f(q_f) - \frac{\beta_m}{2} Q_m^2 - \frac{\beta_m}{2} \sum_{f \in \Fcal(m)}q_f^2\right] \notag \\
& \qquad + \sum_{m\in\Mcal} \left[ (\theta_C - 2\theta_M) \frac{\beta_m}{2} r_m^2 + \theta_M\alpha_m r_m \right].  \label{eq:piHat}
\end{align}
Recall that $\theta_M + \theta_P - \theta_C > 0$. To finish the proof, differentiate the above with respect to $q_f$ to get
\begin{align*}
{\left(\theta_M + \theta_P - \theta_C\right)}^{-1} \nabla_{q_f}\hat{\Pi}(\v{q},\v{r};\v{\theta})
&=  \alpha_{\Mcal(f)} - \beta_{\Mcal(f)} r_{\Mcal(f)}  -  c'_f(q_f) - \beta_{\Mcal(f)} Q_{\Mcal(f)} - {\beta_{\Mcal(f)}} q_f \\
&= \nabla_{q_f} \left[ \{\alpha_{\Mcal(f)} - \beta_{\Mcal(f)} (r_{\Mcal(f)} + Q_{\Mcal(f)})\}q_f - c_f(q_f)\right] \\
&= \nabla_{q_f} \pi_f (\v{q},\v{r}).
\end{align*}
%


\subsection{Proof of Theorem \ref{thm:main}(a).}
\label{sec:proof.main.a}
We tackle the two cases separately.

\newcommand{\Ght}{{\hat{\Gset}(\v{\theta})}}
\paragraph{When $2 \theta_M - \theta_C \geq 0$.}
Rosen's result in Lemma \ref{lemma:generalGames}(a) is used to show that an equilibrium exists. Lemma \ref{lemma:generalGames}(a) does not apply directly to $\Gt$ since its joint strategy set is unbounded. To circumvent that difficulty, we define an auxiliary concave game $\Ght$ with a bounded joint strategy set whose Nash equilibria are also equilibria of $\Gt$, allowing us to leverage Rosen's result.

Recall that $\v{r}$ lies in the compact set $\Pcal'$ defined in \eqref{eq:defP'}. There exists $\bar{r} \in \Rset_+$ such that $\lvert r_m \rvert \leq \bar{r}$ for every $m \in \Mcal$. Define a game $\Ght$ that is identical to $\Gt$ except that the strategy set of each firm $f \in \Fcal$ is restricted to $[0,\bar{q}]$, where
$$\bar{q} := \frac{1}{2}\max_{f \in \Fcal} \left(\alpha_{\Mcal(f)}/\beta_{\Mcal(f)} + \bar{r}\right).$$
Now, $\pi_f$ is continuous in its arguments, and is concave in $q_f$ because
$\nabla^2_{q_f, q_f}\pi_f(\v{q},\v{r})
=
-2\beta_{\Mcal(f)} - c_f^{\prime\prime}(q_f) < 0$. The notation $\nabla^2$ stands for the second (partial) derivative. Then, for $q_f > \bar{q}$, we have
\begin{align}
\nabla_{q_f}\pi_f(q_f, \v{q}_{-f}, \v{r})
&=
\alpha_{\Mcal(f)} - \beta_{\Mcal(f)} r_{\Mcal(f)} - \beta_{\Mcal(f)} Q_{\Mcal(f)} - \beta_{\Mcal(f)} q_f - c_f^\prime(q_f)
\notag \\
&<
\alpha_{\Mcal(f)} + \beta_{\Mcal(f)} \bar{r} - 2\beta_{\Mcal(f)} \bar{q}
\notag \\
&\leq
0. \label{eq:qBound1}
\end{align}
The first inequality in the above chain follows from the fact that $c_f$ is nondecreasing, $|r_{\Mcal(f)}| \leq \bar{r}$, and $Q_{\Mcal(f)} \geq q_f > \bar{q}$. The second one follows from the definition of $\bar{q}$. As a result of the above inequality, $\pi_f(\v{q},\v{r})$ decreases in $q_f$ beyond $\bar{q}$. Now, consider a Nash equilibrium of $\Ght$. Supplier $f$'s action at equilibrium not only maximizes $\pi_f$ over $[0, \bar{q}]$, the above inequality implies that it also does so over $\Rset_+$. Said differently, that equilibrium of $\Ght$ is also an equilibrium of $\Gt$. To conclude the proof using Lemma \ref{lemma:generalGames}(a), we now argue that $\Ght$ is a concave game.

The joint strategy set of $\Ght$ is given by the compact set $[0, \bar{q}]^{| \Fcal |} \times \Pcal'$. The payoff $\pi_f$ of firm $f$ is continuous in all its arguments and has been shown to be concave in $q_f$. The market maker's payoff $\Pi(\v{q},\v{r};\v{\theta})$ is again continuous in all its arguments, and
\begin{align}
\nabla^2_{r_m, r_m'} \Pi (\v{q}, \v{r}; \v{\theta})
& =
\begin{cases} -(2\theta_M - \theta_C)\beta_m, & \text{if } m = m', \\ 0, & \text{otherwise}.\end{cases}
\label{eq:concave}
\end{align}
The nonnegativity of $2\theta_M - \theta_C$ in conjunction with the above relation yields that the Hessian of $\Pi(\v{q},\v{r},\v{\theta})$ with respect to $\v{r}$ is negative semidefinite, implying, $\Pi(\v{q},\v{r};\v{\theta})$ is concave in $\v{r}$. Thus, $\Ght$ is a concave game.


\paragraph{When $\theta_M + \theta_P - \theta_C > 0$.}
Theorem \ref{thm:pot} implies that $\Gt$ is a weighted potential game with $\hat{\Pi}(\v{q},\v{r};\v{\theta})$ as the potential function. We show that the super-level sets of $\hat{\Pi}(\v{q},\v{r};\v{\theta})$ are compact, and hence, $\Ct$ admits a finite optimizer. Lemma \ref{lemma:generalGames}(b) handles the rest. We first appeal to the continuity of $\hat{\Pi}$ to guarantee that the super-level sets are closed. The rest of the proof argues the boundedness of these sets. Utilize \eqref{eq:piHat} to rewrite $\hat{\Pi}$ as
\begin{align*}
\hat{\Pi}(\v{q},\v{r};\v{\theta})
&= \sum_{m \in \Mcal}  [ (\theta_M + \theta_P - \theta_C)  g_m(\v{q}, \v{r}) + h_m(\v{r}) ],
\end{align*}
where $g_m$ and $h_m$ are defined as
\begin{align*}
g_m(\v{q}, \v{r}) &:= (\alpha_m - \beta_m r_m) Q_m - \sum_{f\in\Fcal(m)} c_f(q_f) - \frac{\beta_m}{2} Q_m^2 - \frac{\beta_m}{2} \sum_{f \in \Fcal(m)}q_f^2,\\
h_m(\v{r}) &:= (\theta_C - 2\theta_M) \frac{\beta_m}{2} r_m^2 + \theta_M\alpha_m r_m.
\end{align*}
Recall that $\v{r}$ varies over the compact set $\Pcal'$. Continuity of $h_m$ yields a uniform upper bound on its value as $\v{r}$ varies over $\Pcal'$. Compactness of $\Pcal'$ further implies that there exists $\bar{r} > 0$ such that $| r_m | \leq \bar{r}$ for all $m$. Therefore
$$g_m(\v{q}, \v{r}) \leq (\alpha_m + \beta_m \bar{r}) Q_m - \sum_{f\in\Fcal(m)} c_f(q_f) - \frac{\beta_m}{2} Q_m^2 - \frac{\beta_m}{2} \sum_{f \in \Fcal(m)}q_f^2.$$
The right hand-side of the above relation approaches $-\infty$ as $\vnorm{\v{q}}_2 \to \infty$, where $\vnorm{\cdot}_2$ denotes the $\ell_2$ norm. The inequality implies the same behavior for $\hat{\Pi}(\v{q},\v{r};\v{\theta})$ as well, completing the proof.

\subsection{Proof of Theorem \ref{thm:main}(b) and Corollary \ref{corr:unboundedP}.}
\label{sec:proof.main.b}

Begin with the definition
$$ \gamma^+ := \frac{2\theta_M - \theta_C}{\theta_M + \theta_P - \theta_C} > 0.$$
The potential function $\hat{\Pi}(\v{q}, \v{r}; \v{\theta})$ is twice continuously differentiable. When $\gamma^+ \geq \gamma$, we show that the potential function is jointly concave in its arguments. Then, Lemma \ref{lemma:generalGames}(b) allows us to equate the equilibria of $\Gt$ to the optimizers of $\Ct$. The last step only requires $\Pcal$ to be closed; its boundedness is not necessary. If in addition, $\Pcal$ is bounded, then Theorem \ref{thm:main}(a) guarantees the existence of at least one equilibrium.

If $\gamma^+ > \gamma$, we argue that the potential is \emph{strongly concave}\footnote{A  function $h(\v{x})$ is said to be strongly concave if $h(\v{x}) + \frac{\mu}{2}\vnorm{\v{x}}_2^2$ is concave for some $\mu > 0$.}. For completeness, we show that a twice continuously differentiable strongly concave function always admits a finite maximizer over a closed set. In our context, it implies that a Nash equilibrium always exists and is unique. These observations taken together prove both Theorem \ref{thm:main}(b) and Corollary \ref{corr:unboundedP}.

The following notation will prove useful. Let $\bone$ denote a vector of all ones and $\v{I}$ denote an identity matrix, both of appropriate dimensions. For a symmetric matrix $\v{X}$, let $\v{X} \succeq 0$ (resp. $\v{X} \succ 0$) denote that $\v{X}$ is positive semidefinite (resp. positive definite).

Using this notation, the Hessian of $\hat{\Pi}(\v{q},\v{r};\v{\theta})$ with respect to $(\v{q}, \v{r})$ can be shown to be block diagonal with $| \Mcal |$ block matrices. The block corresponding to market $m\in\Mcal$ is
\begin{align*}
\v{H}_m
=
-
\begin{pmatrix}
\left(\theta_M + \theta_P - \theta_C\right)\left(\beta_m \bone \bone^\top + \diag({\v{d}}_m)\right) & \left(\theta_M + \theta_P - \theta_C\right)\beta_m \bone
\\
\left(\theta_M + \theta_P - \theta_C\right)\beta_m \bone^\top & \left(2\theta_M - \theta_C\right) \beta_m
\end{pmatrix},
\end{align*}
where ${\v{d}}_m := \left( \beta_m + c^{\prime\prime}_f (q_f),  f\in\Fcal(m) \right) \in \Rset^{| \Fcal(m) |}_{+}$. Algebraic operations further yield
\begin{align*}
-\v{H}_m (\mu) := -\frac{1}{\theta_M + \theta_P - \theta_C}\v{H}_m - {\mu} \v{I}
& =
\begin{pmatrix}
\beta_m \bone \bone^\top + \diag({\v{d}}_m)  - \mu \v{I} & \beta_m \bone
\\
\beta_m \bone^\top & \gamma^+ \beta_m - \mu
\end{pmatrix}.
\end{align*}
We now proceed to show:
\begin{itemize}
\item $\gamma^+ \geq \gamma$ implies that $-\v{H}_m(0) \succeq 0$ for each $m$, implying $\hat{\Pi}$ is concave.
\item $\gamma^+ > \gamma$ implies that there exists $0 < \mu < \min_{m \in \Mcal}\{\beta_m\}$ for which $-\v{H}_m(\mu) \succeq 0$ for each $m$, implying $\hat{\Pi}$ is strongly concave.
\end{itemize}
Bearing these objectives in mind, notice that the top-left block of $-\v{H}_m(\mu)$ satisfies
\begin{align*}
\beta_m \bone \bone^\top  + \underbrace{\diag({\v{d}}_m) - \mu \v{I}}_{:= \v{D}_m} \succ 0
\end{align*}
for $0 \leq \mu < \min_{m \in \Mcal}\{\beta_m\}$.
Using Schur complements, the above condition implies that
\begin{align}
-\v{H}_m(\mu) \succeq 0,
& \quad \text{  if  }  \quad
\gamma^+ \beta_m -  \mu  - \beta_m\bone^\top (\beta_m\bone\bone^\top + \v{D}_m)^{-1} \beta_m\bone \geq 0,
\notag \\
& \quad \text{  iff  } \quad
\gamma^+
\geq \frac{\mu}{\beta_m} + \beta_m \bone^\top (\beta_m\bone\bone^\top + \v{D}_m)^{-1} \bone.
\label{eq:HmCriterion}
\end{align}
The Sherman-Morrison-Woodbury matrix identity (see \cite{horn2012matrix}) and elementary algebra yields
\begin{align*}
\beta_m\bone^\top (\beta_m\bone\bone^\top + \v{D}_m)^{-1} \bone
&=
\beta_m\bone^\top \left[\v{D}_m^{-1} - \beta_m \frac{\v{D}_m^{-1} {\bone \bone^\top} \v{D}_m^{-1}}{1 + \beta_m \bone^\top \v{D}_m^{-1} \bone}\right]   \bone \\
&= 1 - \frac{1}{1 + \beta_m \bone^\top \v{D}_m^{-1} \bone} \\
&=
1 - \left({1 + \sum_{f \in \Fcal(m)} \frac{\beta_m}{\beta_m - \mu + c_f^{''} (q_f)} }\right)^{-1}\\
& \leq
1 -\left({1 + \sum_{f \in \Fcal(m)} \frac{\beta_m}{\beta_m - \mu + \inf_{q_f \geq 0} c_f^{''} (q_f)} }\right)^{-1}.
\end{align*}
Plugging the above relation in \eqref{eq:HmCriterion}, we obtain
\begin{align*}
-\v{H}_m(\mu) \succeq 0, \quad \text{ if } \quad
\gamma^+  \geq   \frac{\mu}{\beta_m} +  1 -\left({1 + \sum_{f \in \Fcal(m)} \frac{\beta_m}{\beta_m - \mu + \inf_{q_f \geq 0} c_f^{''} (q_f)} }\right)^{-1}.
\end{align*}
The right-hand side is continuous at $\mu = 0$ for each $m$. Their maximum over $\Mcal$ is  also continuous at $\mu = 0$, where it equals $\gamma$. Taken together, it implies that $\hat{\Pi}$ is concave (resp. strong concave) when $\gamma^+ \geq \gamma$ (resp. $\gamma^+ > \gamma$). We now state and prove the following technical result on strongly concave functions that implies the rest.

\begin{lemma}
\label{lemma:strongConcave}
If a twice continuously differentiable scalar-valued function $h$ is strongly concave, i.e., $h(\v{x}) + \frac{\mu}{2}\vnorm{\v{x}}_2^2$ is concave for some $\mu > 0$, then $h$ has a unique finite maximizer.
\end{lemma}
\proof{Proof.}
Smoothness and concavity of $h(\v{x}) + \frac{\mu}{2}\vnorm{\v{x}}_2^2$ together imply that $-\nabla^2 h - \mu \v{I} \succeq 0$ everywhere. Taylor's expansion of $h$ and Cauchy-Schwarz inequality then imply
$$ h(\v{x})  \leq h(\v{y}) + \langle \nabla h(\v{y}), \v{x} - \v{y} \rangle - \mu \vnorm{\v{x} - \v{y}}_2^2 \leq h(\v{y}) + \vnorm{ \nabla h(\v{y})}_2 \cdot \vnorm{\v{x} - \v{y}}_2  - \mu \vnorm{\v{x} - \v{y}}_2^2$$
for arbitrary $\v{x}$ and $\v{y}$ in the domain of $h$. For a fixed $\v{y}$, the right-hand side approaches $-\infty$ as $\vnorm{\v{x} - \v{y}}_2 \to \infty$. Owing to the above inequality, so does $h(\v{x})$. Thus, the super-level sets of $h$ are bounded; continuity of $h$ guarantees that all maximizers lie in that bounded set. Uniqueness follows from strict concavity of $h$ that is implied by its strong concavity.
\Halmos
\endproof

\subsection{Proof of Proposition \ref{prop:Pfull}.}
\label{sec:proof.Pfull}
The conditions imposed on $\Gcal^u$ satisfy the requirements of Corollary \ref{corr:unboundedP}. Thus, the unique Nash equilibrium of $\Gcal^u$ is given by the unique optimizer of $\Ct$. Notice that $\Ct$ in this setting is equivalent to solving a convex optimization problem. Then, Karush-Kuhn-Tucker (KKT) optimality conditions are both necessary and sufficient. Unpacking these conditions, we get that $(\v{q}, \v{r})$ solves $\Ct$ if and only if $\v{q}\in\Rset_+^{|\Fcal|}$, $\bone^\top\v{r} = 0$, and there exists $\v{\mu}\in\Rset_+^{|\Fcal|}$ and $\lambda\in\Rset$ that satisfy
\begin{align}
\label{eq:PfullKKT}
\nabla_{\v{r}} [ \hat{\Pi}(\v{q}, \v{r}; \v{\theta}) - \lambda \bone^\top \v{r}  ]=\v{0}, \qquad
\nabla_{\v{q}} [ \hat{\Pi}(\v{q}, \v{r}; \v{\theta}) + \v{\mu}^\top \v{q}] = \v{0}, \qquad
\v{\mu}^\top \v{q} & = 0.
\end{align}
%
We argue that $\v{q},\v{r}$ in \eqref{eq:Pfull.q} -- \eqref{eq:Pfull.r}, and $\v{\mu}, \lambda$ defined below together satisfy the optimality conditions.
\begin{equation}
\label{eq:defLambda}
\v{\mu} = \mathbf{0}, \qquad \lambda := \frac{1}{2}\left(\theta_P +\theta_M-\theta_C\right) \left( \bar{C} - \alpha \right) + \theta_M \alpha.
\end{equation}
The lower bound on $\alpha$ implies that $\v{q}$ is elementwise nonnegative, and
$ \bone^\top\v{r} = (\kappa(\vt)/\beta) \sum_{f \in \Fcal} (C_f - \bar{C}) = 0$.
The relations in \eqref{eq:PfullKKT} take the form
\begin{align*}
\nabla_{r_{\Mcal(f)}}[\hat\Pi(\v{q},\v{r};\v{\theta}) - \lambda\bone^\top\v{r}]
&=
- \left( 2\theta_M - \theta_C\right)\beta r_{\Mcal(f)} - \left(\theta_P+\theta_M-\theta_C\right)\beta q_f + \theta_M \alpha - \lambda
=
0, \\
\nabla_{q_f}
[\hat{\Pi}(\v{q},\v{r};\v{\theta}) + \v{\mu}^\top\v{q}]
&=
\alpha - \beta \left( r_{\Mcal(f)} + 2 q_f\right) - C_f + \mu_f
=0, \\
\v{\mu}^\top \v{q} &= 0.
\end{align*}
The first among the above relations require the definition of $\lambda$ from \eqref{eq:defLambda}.
In summary, we have verified the optimality conditions and conclude that $(\v{q},\v{r})$ in \eqref{eq:Pfull.q} -- \eqref{eq:Pfull.r} is the unique equilibrium of $\Gcal^u$. After some algebra, we further obtain
\begin{align*}
\sum_{m\in\Mcal}\mathsf{CS}_m(\v{q},\v{r})
&=
\frac{\beta}{2}\sum_{f\in\Fcal}  \left(q_f + r_{\Mcal(f)}\right)^2
\\
&=
\frac{1}{8\beta}\left[ |\Fcal|  (\alpha - \bar{C})^2 + \left(\kappa(\v{\theta})-1\right)^2 \sum_{f\in\Fcal}(C_f - \bar{C})^2 \right];
\\
\sum_{m\in\Mcal}\mathsf{PS}_m(\v{q},\v{r})
&=
\sum_{f\in\Fcal}\left[\alpha - C_f - \beta \left(q_f + r_{\Mcal(f)}\right)\right] \cdot q_f
\\
&=
\frac{1}{4\beta} \left[ |\Fcal| (\alpha - \bar{C})^2 + \left( \kappa(\v{\theta}) + 1\right)^2 \sum_{f\in\Fcal}(C_f - \bar{C})^2\right];
\\
\sum_{m\in\Mcal}\mathsf{MS}_m(\v{q},\v{r})
&=
-\beta \sum_{f\in\Fcal} \left(q_f + r_{\Mcal(f)}\right) \cdot r_{\Mcal(f)}
\\
&=
-\frac{1}{2\beta}\kappa(\v{\theta})(\kappa(\v{\theta})-1)\sum_{f\in\Fcal}(C_f - \bar{C})^2.
\end{align*}
Then, the social welfare at the unique Nash equilibrium is given by
\begin{align*}
&\sum_{m\in\Mcal}\left[\mathsf{CS}_m(\v{q},\v{r}) + \mathsf{PS}_m(\v{q},\v{r}) + \mathsf{MS}_m(\v{q},\v{r})\right]
\\
& \quad =
\frac{3|\Fcal|}{8\beta}\left(\alpha - \bar{C}\right)^2 + \frac{1}{8\beta}\left[(\kappa(\v{\theta})-1)^2 + 2(\kappa(\v{\theta})+1)^2 - 4\kappa(\v{\theta})(\kappa(\v{\theta})-1)\right] \sum_{f\in\Fcal}\left(C_f - \bar{C}\right)^2
\\
& \quad =
\frac{3|\Fcal|}{8\beta}\left(\alpha - \bar{C}\right)^2 + \frac{1}{8\beta}\left(-\kappa(\v{\theta})^2 + 6\kappa(\v{\theta}) + 3\right) \sum_{f\in\Fcal}\left(C_f - \bar{C}\right)^2.
\end{align*}
The definition of $\sigma_c^2$ in the above equation yields the desired result.


\subsection{Reformulating \eqref{eq:findTheta.2} as a polynomial optimization problem.}
\label{sec:proof.SOS}

We argue that $\Ct$ can be reformulated in a way that the search for its primal-dual optimizers over $\Theta_\ve$ can be restricted to a bounded set. The rest follows from rewriting the KKT equations as polynomial inequalities.

Compactness of $\Pcal'$ implies that there exists $\bar{r} \geq | r_m |$ for each component of $\v{r} \in\Pcal'$. Recall from \eqref{eq:qBound1} that $\pi_f$ (and hence $\hat{\Pi}$) decreases in $q_f$ beyond $\bar{q} := \frac{1}{2}( \max_{m \in \Mcal} \alpha_m/\beta_m + \bar{r})$. Then, it is enough to search for $q_f$ over $[0, \bar{q}]$ for an optimizer of $\Ct$.

$\Pcal'$ is not full dimensional. Project it to define it as a linear map of a full dimensional polytope $\hat{\Pcal}$ that is described by $\hat{\v{A}} \hat{\v{r}} \leq \hat{\v{b}}$. We elaborate on this step towards the end of the proof. Collect $\v{q}$ and $\hat{\v{r}}$ in $\v{x}$ and write
$$ \Ct \  \equiv \ \text{maximize} \ \ \hat{\Pi}(\v{x}; \vt), \text{ subject to } \v{D} \v{x} \leq \v{d}.$$
The constraint $\v{D} \v{x} \leq \v{d}$ encodes $\mathbf{0} \leq \v{q} \leq \bar{q}\bone$ and $\hat{\v{A}} \hat{\v{r}} \leq \hat{\v{b}}$. Call its finite optimal value $\hat{\Pi}^*(\vt)$.

Critical to our proof is the observation that there exists a strictly feasible point $\bar{\v{x}}$, i.e., $ \v{D} \bar{\v{x}} < \v{d}$. Therefore, Slater's condition holds. Consequently, $\Ct$ satisfies strong duality and the dual optimal value is attained. Let $\v{\lambda}(\vt) \geq 0$ be an optimal Lagrange multiplier for the inequality constraint. We argue that its  $\ell_1$-norm $\vnorm{\lambda(\vt)}_1$ admits a uniform upper bound over $\Theta_\ve$ using an argument that mimics \cite[Lemma 3]{nedic2009subgradient}. Denote the minimum among the elementwise positive vector $\v{d} - \v{D} \bar{x}$ as $\bar{d} > 0$. Then, we have
\begin{align*}
\hat{\Pi}^*(\vt)
& = \max_{\v{x} : \v{D} \v{x} \leq \v{d}} \left[  \hat{\Pi}({\v{x}}; \vt) \ + \ \langle \v{\lambda}(\vt),  \v{d} - \v{D} {\v{x}}  \rangle \right] \\
& \geq \hat{\Pi}(\bar{\v{x}}; \vt) \ + \ \langle \v{\lambda}(\vt),  \v{d} - \v{D} \bar{\v{x}} \rangle \\
& \geq \hat{\Pi}(\bar{\v{x}}; \vt) + \vnorm{\v{\lambda}(\vt)}_1 \bar{d}.
\end{align*}
Here, $\langle \cdot, \cdot \rangle$ denotes the usual dot product. The first inequality is derived from the optimality of $\hat{\Pi}^*(\vt)$ and the second one follows from the definition of $\bar{d}$. A simple rearrangement yields
$$ \vnorm{\v{\lambda}(\vt)}_1 \leq \frac{1}{\bar{d}} \left( \hat{\Pi}^*(\vt) -  \hat{\Pi}(\bar{\v{x}}; \vt)  \right). $$
Notice that $\hat{\Pi}$ has bounded variation over $\v{D} \v{x} \leq \v{d}$ for all $\vt \in \Theta_\ve$. Combined with the above inequality, that provides a uniform upper bound on $\vnorm{\v{\lambda}(\vt)}_1$ over $\Theta_\ve$.

For completeness, we outline the procedure to obtain $\hat{\Pcal}$ from $\Pcal'$. Distinguish all implicit equalities and remove all redundancy in the description of $\Pcal'$ to write $\Pcal'$ as
$\{ \v{r} :  \v{A}_e \v{r} = \v{b}_e, \ \v{A}_i \v{r} \leq \v{b}_i\}.$\footnote{One can use linear programming to detect redundancy and implicit equalities in linear inequality systems. See \cite{greenberg1996consistency} and \cite[Chapter 8]{schrijver1998theory} for a discussion on the topic.
} Then, $\v{A}_e$ has full row-rank (call it $n$), and $\Pcal'$ is an $|\Mcal| - n$ dimensional polytope. Rearrange the columns of $\v{A}_e$ (and rows of $\v{r}$) so that the first $n$ columns of $\v{A}_e$ are linearly independent. Distinguish them as $\v{A}_e = ( \v{A}_e^n \ \v{A}'_e)$. Conformally partition $\v{A}_i$ as $( \v{A}_i^n \ \v{A}'_i)$. Use the invertibility of $\v{A}_e^n$ to obtain
\begin{align*}
\hat{\Pcal} &:= \left\{ \hat{\v{r}} \in \Rset^{|\Mcal| - n} : \left(\v{A}'_i - \v{A}_i^n [\v{A}_e^n]^{-1} \v{A}'_e \right) \hat{\v{r}} \leq \v{b}_i - \v{A}_i^n [\v{A}_e^n]^{-1} \v{b}_e \right\}, \\
\Pcal' &=
\left\{
\begin{pmatrix}
[\v{A}_e^n]^{-1} \left( \v{b}_e -  \v{A}'_e \hat{\v{r}} \right)
\\
\hat{\v{r}}
\end{pmatrix} : \hat{\v{r}} \in \hat{\Pcal}
\right\}.
\end{align*}
$\hat{\Pcal}$ is full-dimensional and always contains a strictly feasible point (see \cite[Section 8.1]{schrijver1998theory}).

\section{Analyzing the Two-Market Two-Firm Example in Fig. \ref{fig:2Node}}
\label{sec:two-node}

In this section, we derive all Nash equilibria of $\Gt$ over $\Delta$ in a two-market two-firm example, portrayed in Figure \ref{fig:2Node}. The formulae allow us to draw insights into the effect of the design parameter on the nature of the equilibria.

Suppose each firm has an increasing linear cost, given by $c_f(q_f) = C_f q_f$, and the markets have identical demand functions described by $p_m(d_m) = \alpha - \beta d_m$ for $m=1,2$.  The nodal markets are joined via a link of capacity $b$, modeled as $\Pcal := \{ \v{r} = (r_1, r_2)^\top : | r_1 | \leq b, \ |r_2| \leq b\}$. In essence, the two nodal markets only differ in the marginal costs of the firms supplying in them.

The market maker neither consumes nor supplies. Use $r := r_1 = -r_2$ to simplify its strategy set to $\{ r\in\Rset : |r| \leq b \}$. Restrict attention to the case $\alpha \geq b \beta + \max\{C_1, C_2\}$. In that setting, the optimal response of the firms is given by
\begin{align}
q_1 = \frac{1}{2}\left(\frac{\alpha-C_1}{\beta} - r\right), 
\quad
q_2 = \frac{1}{2}\left(\frac{\alpha-C_2}{\beta} + r\right).
\label{eq:qBR}
\end{align}
The firms' productions defined above, together with $r$, constitute an equilibrium if $r$ maximizes 
\begin{align*}
\Pi(q_1,q_2,r;\vt)
&=
-\left(\theta_M+\theta_P-\theta_C\right)\left(q_1 - q_2\right)\beta r - \left(2\theta_M-\theta_C\right)\beta r^2 \\
& \qquad + \theta_P\left(\left(\alpha-C_1\right)q_1+\left(\alpha-C_2\right)q_2\right) + \frac{1}{2}\left(\theta_C-2\theta_P\right)\beta\left(q_1^2+q_2^2\right)
\end{align*}
over $[-b, b]$. We characterize $\Rcal(\vt)$, the set of all $r$'s in an equilibrium of $\Gt$ over $\Delta$ in Lemmas \ref{lem:equilibrium-part1} -- \ref{lem:equilibrium-part3}. The findings are summarized in Table \ref{tbl:2node-summary}. Our results make use of the following additional notation. For any $x\in\Rset$, let ${[x]}_\ell^u$ denote the projection of $x$ on the interval $[\ell, u]$, and $\sgn(x)$ denote its sign. Finally, $\varnothing$ stands for a null set, and define
\begin{align*}
\Delta C := C_1 - C_2, \quad \text{and}\quad\kappa(\vt) :=
\frac{\theta_P + \theta_M - \theta_C}{3\theta_M - \theta_C - \theta_P}.
\end{align*}

\begin{table}[h!]
\centering
\small
{\begin{tabular}{| l | l | l |}
\hline
\multicolumn{2}{|c|}{Conditions on $\vt$}
&
\multicolumn{1}{c|}{$\Rcal(\vt)$}
\\
\hline
\hline
\multirow{5}{*}{$2\theta_M - \theta_C > 0$} 
& $3\theta_M-\theta_C-\theta_P > 0$ 
& $\left\{\left[\frac{\kappa(\vt) \Delta C}{2\beta}\right]_{-b}^{+b}\right\}$
\\
\cline{2-3}
& $3\theta_M-\theta_C-\theta_P = 0$ 
& $\begin{aligned}
&\left[-b,+b\right], && \text{if} \ \Delta C = 0,
\\
&\left\{b \cdot \sgn(\Delta C)\right\}, && \text{otherwise}.
\end{aligned}$
\\
\cline{2-3}
& $3\theta_M-\theta_C-\theta_P < 0$ 
& $\begin{aligned}
&\left\{\pm b, \frac{\kappa(\vt) \Delta C}{2\beta}\right\}, && \text{if} \ \left|\frac{\kappa(\vt) \Delta C}{2\beta}\right| \leq b,
\\
&\left\{b \cdot \sgn(\Delta C)\right\}, && \text{otherwise}.
\end{aligned}$
\\
\hline
\multirow{4}{*}{$2\theta_M-\theta_C = 0$} 
& $\theta_M+\theta_P-\theta_C < 0$ 
& $\left\{\left[-\frac{\Delta C}{2\beta}\right]_{-b}^{+b}\right\}$
\\
\cline{2-3}
& $\theta_M+\theta_P-\theta_C = 0$ 
& $\left[-b,+b\right]$
\\
\cline{2-3}
& $\theta_M+\theta_P-\theta_C > 0$ 
& $\begin{aligned}
& \left\{\pm b, -\frac{\Delta C}{2\beta}\right\}, && \text{if} \ \left|\frac{\Delta C}{2\beta}\right| \leq b,
\\
& \left\{b \cdot \sgn(\Delta C)\right\}, && \text{otherwise}.
\end{aligned}$
\\
\hline
\multirow{5}{*}{$2\theta_M-\theta_C < 0$} 
& $\theta_M+\theta_P-\theta_C < 0$ 
& $\begin{aligned}
&\left\{-b \cdot \sgn(\Delta C)\right\}, && \text{if} \ \left|\frac{\Delta C}{2\beta}\right| \geq b,
\\
&\varnothing, && \text{otherwise}.
\end{aligned}$
\\
\cline{2-3}
& $\theta_M+\theta_P-\theta_C = 0$ 
& $\left\{\pm b\right\}$
\\
\cline{2-3}
& $\theta_M+\theta_P-\theta_C > 0$ 
& $\begin{aligned}
&\left\{\pm b\right\}, && \text{if} \ \left|\frac{\Delta C}{2\beta}\right| \leq b,
\\
&\left\{b \cdot \sgn(\Delta C)\right\}, && \text{otherwise}.
\end{aligned}$
\\
\hline
\end{tabular}}
\centering
\caption{Summary of $\Rcal(\vt)$ for the two-market two-firm example. The equilibria of $\Gt$ are given by $(q_1, q_2, r)$, where $r \in \Rcal(\vt)$ and $q_1$ and $q_2$ are defined through \eqref{eq:qBR}.
\label{tbl:2node-summary}}
\end{table}

Theorems \ref{thm:pot} and \ref{thm:main} only provide sufficient conditions for $\Gt$ to exhibit certain properties. While we do not provide any tightness results, Table \ref{tbl:2node-summary} helps to demonstrate that each property may fail to hold unless the conditions in the theorems are satisfied. Some highlights from the analysis:

\begin{enumerate}

\item When neither $\theta_M + \theta_P - \theta_C > 0$ nor $2\theta_M - \theta_C \geq 0$ holds, an equilibrium may not exist. An example is a $\vt$ where each of the above quantities are negative and $\left|\frac{\Delta C}{2\beta}\right| < b$. 

\item When $2\theta_M - \theta_C \geq 0$, but not $\theta_M+\theta_P - \theta_C > 0$, an equilibrium exists, but $\Gt$ is not a potential game.
Consider $\vt$ with $2\theta_M - \theta_C = 0$, $\theta_M+\theta_P-\theta_C < 0$ and $C := C_1 = C_2$. Table \ref{tbl:2node-summary} suggests a unique equilibrium with $\Rcal(\vt) = \{0 \}$. Dynamics of $\Mcal$ and firms 1, 2 sequentially playing best response to others starting from $r = +b, q_1= \frac{1}{2}\left(\frac{\alpha-C}{\beta} - b\right)$ and $q_2 = \frac{1}{2}\left(\frac{\alpha-C}{\beta} + b\right)$ results in
\begin{align*}
& r = -b 
\ \to \
q_1 = \frac{1}{2}\left(\frac{\alpha-C}{\beta} + b\right) 
\ \to \
q_2 = \frac{1}{2}\left(\frac{\alpha + C}{\beta} - b\right) \\
& \quad 
\to \
r = +b 
\ \to \
q_1 = \frac{1}{2}\left(\frac{\alpha-C}{\beta} - b\right) 
\ \to \
q_2 = \frac{1}{2}\left(\frac{\alpha + C}{\beta} + b\right), 
\end{align*}
confirming the presence of a cycle, precluding $\Gt$ from being a potential game (not just with our candidate potential function).

%

\item When $2\theta_M - \theta_C < 0$ and $\theta_M+\theta_P-\theta_C > 0$, equilibrium exists despite the possible loss of concavity in the market maker's objective. Table \ref{tbl:2node-summary} corroborates that conclusion in our example.

\item When $2\theta_M - \theta_C \geq 0$, $\theta_M+\theta_P-\theta_C > 0$, but not $2\theta_M - \theta_C \geq \gamma \cdot \left(\theta_M+\theta_P-\theta_C\right)$, not all equilibria are optimizers of the potential function. In our example,
$$ \gamma = \frac{1}{2} \ \implies \ (2\theta_M - \theta_C) -  \gamma \cdot \left(\theta_M+\theta_P-\theta_C\right) = \frac{1}{2}(3\theta_M-\theta_C - \theta_P).$$
With $2\theta_M - \theta_C > 0$, $3\theta_M-\theta_C - \theta_P < 0$, and $\left|\frac{\kappa(\vt) \Delta C}{2\beta}\right| < b$, Table \ref{tbl:2node-summary} reveals three distinct equilibria with $r = \pm b, \frac{\kappa(\vt) \Delta C}{2\beta}$. 
Any optimizer of $\Ct$ satisfies \eqref{eq:qBR}, using which $\hat{\Pi}$ becomes
\begin{align*}
\frac{1}{2}(\theta_M+\theta_P-\theta_C)\left[ r \cdot \Delta C + \frac{1}{2\beta}(\alpha - C_1)^2 + \frac{1}{2\beta}(\alpha - C_2)^2\right] - \frac{1}{2}\left(3\theta_M - \theta_C - \theta_P\right) \beta r^2.
\end{align*}
The above is strictly convex in $r$ and $\pm b$ are the only candidate optimizers. Said, otherwise, the equilibrium with $r = \frac{\kappa(\vt) \Delta C}{2\beta}$ is not an optimizer of $\Ct$. 

\item When $2\theta_M - \theta_C = \gamma \left(\theta_M+\theta_P-\theta_C \right) > 0$, then $\Gt$ may have multiple equilibria, all of which are optimizers of $\Ct$. This is observed in our example, where $\Rcal(\vt) = [-b,+b]$ with $2\theta_M - \theta_C > 0$, $3\theta_M-\theta_C-\theta_P = 0$, and $C_1 = C_2$. 

\end{enumerate}

\subsection{Characterizing the equilibria for the example}
\label{sec:exampleChar}
We state and prove a sequence of lemmas that define $\Rcal(\vt)$ given in Table \ref{tbl:2node-summary}.

\begin{lemma}
\label{lem:equilibrium-part1}
Suppose $\theta_M+\theta_P-\theta_C = 0$. Then, $\Rcal(\vt)$ is given by
\begin{align*}
\Rcal(\v{\theta})=
\begin{cases}
\{0\}, & \text{if } 2\theta_M-\theta_C > 0, \\
\left[-b,+b\right], & \text{if } 2\theta_M-\theta_C = 0, \\
\{\pm b\}, & \text{otherwise}.
\end{cases}
\end{align*}
\end{lemma}
\proof{Proof.}
The maximizer of $\Pi(q_1,q_2,r;\vt)$ over $r$ is independent of $q_1$ and $q_2$. Further, if $2\theta_M-\theta_C > 0$, then $\Pi$ is a concave quadratic even function of $r$. Hence, $r = 0$ is its unique maximizer. On the other hand, if $2\theta_M-\theta_C = 0$, then $\Pi$ is independent of $r$, implying each $r \in \left[-b,b\right]$ constitutes a maximizer of $\Pi$. Finally, if $2\theta_M-\theta_C < 0$, then $\Pi$ is a convex quadratic even function of $r$ attaining its maximum at $r = \pm b$. 
\Halmos
\endproof

\begin{lemma}
\label{lem:equilibrium-part2}
Suppose $\theta_M+\theta_P-\theta_C < 0$. Then, $\Rcal (\vt)$ is given by
\begin{align*}
\Rcal(\v{\theta})=
\begin{cases}
\left\{\left[\frac{\kappa(\vt) \Delta C}{2\beta}\right]_{-b}^{+b}\right\}, & \text{if } 2\theta_M-\theta_C > 0, \\
\left\{\left[-\frac{\Delta C}{2\beta}\right]_{-b}^{+b}\right\}, & \text{if } 2\theta_M-\theta_C = 0, \\
\left\{-b \cdot \sgn(\Delta C)\right\}, & \text{if } 2\theta_M-\theta_C < 0, \text{ and } \left| \frac{\Delta C}{2\beta} \right|  \geq b, \\
\varnothing, & \text{otherwise}.
\end{cases}
\end{align*}
\end{lemma}
\proof{Proof.}
We tackle different cases based on the sign of $2\theta_M-\theta_C$. 
For convenience, call the expressions in \eqref{eq:qBR} as $q_1(r)$ and $q_2(r)$, respectively. The following additional notation will prove useful. The derivative of $\Pi(q_1,q_2,r;\vt)$ with respect to $r$ evaluated at $(q_1(r), q_2(r), r)$ is given by
\begin{align}
\rho(r,\vt) 
& :=
\frac{\Delta C}{2}(\theta_M+\theta_P-\theta_C) - (3\theta_M-\theta_C-\theta_P)\beta r.
\label{eq:2node-rho-theta}
\end{align}

\begin{itemize}
\item When $2\theta_M-\theta_C > 0$, the function $\Pi(q_1,q_2,r;\vt)$ is strictly concave in $r$. Then, the triple $(q_1(r), q_2(r), r)$ constitutes an equilibrium if and only if one of the following three cases arise. 
\begin{align}
\{ | r| \leq b , \ \rho(r,\vt)  = 0 \}, \quad \text{or} \quad 
\{ r = -b, \ \rho(r,\vt)  \leq 0 \}, \quad \text{or} \quad 
\{ r = +b, \ \rho(r,\vt) \geq 0 \}.
\label{eq:part2a-conditions}
\end{align}
Now, $\theta_M+\theta_P-\theta_C < 0$ and $2\theta_M-\theta_C > 0$ imply $3\theta_M-\theta_C-\theta_P > 0$. Then, \eqref{eq:2node-rho-theta} and \eqref{eq:part2a-conditions} yield
$$\Rcal(\vt) = \left\{\left[\frac{\kappa(\vt) \Delta C}{2\beta}\right]_{-b}^{+b}\right\},$$

\item When $2\theta_M-\theta_C = 0$, $\Pi(q_1,q_2,r;\vt)$ is linear in $r$ with slope $\propto (q_1 - q_2)$. Again, an equilibrium of the game can arise in one of three ways: (i) $|r| \leq b$ and $q_1(r) = q_2(r)$, or (ii) $r = +b$ and $q_1(r) >  q_2(r)$, or (iii) $r = -b$ and $q_1(r) <  q_2(r)$. Expanding these conditions using \eqref{eq:qBR}, we get
$$ \Rcal(\vt) = \left\{ \left[ \frac{\Delta C}{2 \beta} \right]^{+b}_{-b} \right\}.$$

\item Finally, when $2\theta_M-\theta_C < 0$, $\Pi(q_1,q_2,r;\vt)$ is strictly convex in $r$ that is maximized either at $r = -b$ or $r = +b$ or at both. 
Notice that
\begin{align*}
\Pi(q_1,q_2,+b;\vt) - \Pi(q_1,q_2,-b;\vt)
&=
-2(\theta_M+\theta_P-\theta_C)(q_1-q_2)\beta b,
\end{align*}
implying $+b$ is an optimizer if $q_1(+b) \geq q_2(+b)$, and $-b$ is an optimizer if $q_1(-b) \leq  q_2(-b)$. Upon simplifying, we get
\begin{equation*}
\Rcal(\v{\theta}) 
= 
\begin{cases}
\left\{-b \cdot \sgn(\Delta C)\right\}, & \text{if} \left| \frac{\Delta C}{2\beta} \right|  \geq b,
\\
\varnothing, & \text{otherwise}.
\end{cases}
\end{equation*}
\end{itemize}
\Halmos
\endproof


\begin{lemma}
\label{lem:equilibrium-part3}
Suppose $\theta_M+\theta_P-\theta_C > 0$. Then, $\Rcal(\vt)$ is given by 
\begin{align*}
\Rcal(\vt) = \begin{cases}
\left\{\left[\frac{\kappa(\vt) \Delta C}{2\beta}\right]_{-b}^{+b}\right\}, 
& \text{if } 2\theta_M-\theta_C > 0, \text{ and } 3\theta_M-\theta_C-\theta_P > 0, \\
\left[-b,+b\right], 
& \text{if } 2\theta_M-\theta_C > 0,  3\theta_M-\theta_C-\theta_P = 0, \text{ and } \Delta C = 0,\\
\left\{\pm b, \frac{\kappa(\vt) \Delta C}{2\beta}\right\}, 
& \text{if } 2\theta_M-\theta_C > 0,  3\theta_M-\theta_C-\theta_P < 0, \text{ and } \left|  \frac{\kappa(\vt) \Delta C}{2\beta} \right| \leq b, \\
\left\{\pm b, -\frac{\Delta C}{2\beta}\right\}, 
& \text{if } 2\theta_M-\theta_C = 0, \text{ and } \left| \frac{\Delta C}{2\beta} \right| \leq b, \\
\left\{\pm b\right\}, 
& \text{if } 2\theta_M-\theta_C < 0, \text{ and } \left| \frac{\Delta C}{2\beta} \right| \leq b, \\
\left\{b \cdot \sgn(\Delta C) \right\}, 
& \text{otherwise}.
\end{cases}
\end{align*}
\end{lemma}
\proof{Proof.}
We divide the analysis based on the sign of $2\theta_M-\theta_C$. First, consider the case with $2\theta_M-\theta_C > 0$. Then, $\Pi(q_1,q_2,r;\vt)$ is strictly concave in $r$. Following our analysis for Lemma \ref{lem:equilibrium-part2}, we infer that $(q_1(r), q_2(r), r)$ becomes an equilibrium in three ways, given by \eqref{eq:part2a-conditions}. We split this case further based on the sign of $3\theta_M-\theta_C-\theta_P $.

\begin{itemize} 

\item When $3\theta_M-\theta_C-\theta_P > 0$: The analysis is identical to the case when $\theta_P + \theta_M -\theta_C < 0$ and $3\theta_M-\theta_C-\theta_P > 0$ that yields 
\begin{align*}
\Rcal(\vt) = \left\{\left[\frac{\kappa(\vt) \Delta C}{2\beta}\right]_{-b}^{+b}\right\}.
\end{align*}

\item When $3\theta_M-\theta_C-\theta_P = 0$: The signs of $\rho(r,\vt)$ from \eqref{eq:2node-rho-theta} matches that of $\Delta C$. Then, \eqref{eq:part2a-conditions} implies  
\begin{align*}
\Rcal(\vt) = \begin{cases}
[-b, +b], & \text{if } \Delta C = 0, \\
\{ b \cdot \sgn(\Delta C), & \text{otherwise}.
\end{cases}
\end{align*}

\item When $3\theta_M-\theta_C-\theta_P < 0$: Here $\rho(r,\vt)$ is increasing in $r$, and \eqref{eq:part2a-conditions} yields
\begin{align*}
\Rcal(\vt)
=\begin{cases}
\left\{\pm b, \frac{\kappa(\vt) \Delta C}{2\beta}\right\}, & \text{if } \left| \frac{\kappa(\vt) \Delta C}{2\beta} \right| \leq b, \\
\left\{b \cdot \sgn(\Delta C) \right\}, & \text{otherwise}.
\end{cases}
\end{align*}
\end{itemize}

Next, turn to the case when $2\theta_M-\theta_C = 0$, where $\Pi(q_1,q_2,r;\vt)$ is linear in $r$ with slope $\propto (q_1 - q_2)$. The analysis is similar to the case when $2\theta_M-\theta_C = 0$, but with $\theta_P + \theta_M - \theta_C <0$. Proceeding as in the proof of Lemma \ref{lem:equilibrium-part2}, we obtain
\begin{equation*}
\Rcal(\vt)
=
\begin{cases}
\left\{\pm b, -\frac{\Delta C}{2\beta}\right\}, & \text{if} \left| \frac{\Delta C}{2\beta} \right| \leq b,
\\
\left\{ b \cdot \sgn(\Delta C) \right\}, & \text{otherwise}.
\end{cases}
\end{equation*}

Finally, when $2\theta_M-\theta_C < 0$, $\Pi(q_1,q_2,r;\vt)$ is strictly convex in $r$ that is maximized at either $r = -b$ or $r = +b$ or both. Again, the analysis mirrors that for Lemma \ref{lem:equilibrium-part2}, and yields
\begin{equation*}
\Rcal(\vt)
=
\begin{cases}
\left\{\pm b\right\}, & \text{if} \left| \frac{\Delta C}{2\beta} \right| \leq b,
\\
\left\{b \cdot \sgn(\Delta C)\right\}, & \text{otherwise}.
\end{cases}
\end{equation*}

\Halmos
\endproof

\bibliographystyle{plain}
\bibliography{bibfile}

\end{document}